\documentclass[conference]{IEEEtran}
\IEEEoverridecommandlockouts
\usepackage{cite}
\usepackage{amsmath,amssymb,amsfonts}
\usepackage{amsthm}
\usepackage{graphicx}
\usepackage{textcomp}
\usepackage{xcolor}
\usepackage{ tipa }
\usepackage{subcaption}
\usepackage{graphicx}
\usepackage{bbm}
\usepackage{verbatim}
\usepackage{mathtools,amssymb}
\usepackage{url}
\usepackage{hyperref}
\usepackage{xcolor}
\usepackage[ruled,vlined]{algorithm2e}
\def\BibTeX{{\rm B\kern-.05em{\sc i\kern-.025em b}\kern-.08em
    T\kern-.1667em\lower.7ex\hbox{E}\kern-.125emX}}
\begin{document}
\SetKwInput{KwInput}{Input}                % Set the Input
\SetKwInput{KwOutput}{Output}              % set the Output
\newcommand{\fairnesscriteria}{\textit{MANI-Rank }}
\newcommand{\fairnesscriterianospace}{\textit{MANI-Rank}}
\newcommand{\exactmethod}{MFRA-IP }
\newcommand{\approxmethod}{MFRA-PC }
\newcommand{\makefairalgo}{Make Mani-Rank}
\newcommand{\fairKemeny}{Fair-Kemeny}
\newcommand{\fairBorda}{Fair-Borda}
\newcommand{\fairSchulze}{Fair-Schulze}
\newcommand{\fairCopeland}{Fair-Copeland}
\newcommand{\fairKemenyspace}{Fair-Kemeny }
\newcommand{\fairBordaspace}{Fair-Borda }
\newcommand{\fairSchulzespace}{Fair-Schulze }
\newcommand{\fairCopelandspace}{Fair-Copeland }
\newcommand{\ourproblemnospace}{MFCR}
\newcommand{\ourproblem}{MFCR }
\newcommand{\mfracritfair}{\ourproblem group-fair criteria}
\newcommand{\mfracritbaser}{\ourproblem pref criteria}
\newcommand{\mfracritfairspace}{\ourproblem group-fair criteria }
\newcommand{\mfracritbaserspace}{\ourproblem pref criteria }
\newcommand{\makefair}{Make-MR-Fair}
\newcommand{\makefairspace}{Make-MR-Fair }
\newcommand{\informalfairproperty}{\textit{Group Fairness}}
\newcommand{\informalconsensusproperty}{\textit{Representation of Preferences}}
\newcommand{\informalfairpropertyspace}{\textit{Group Fairness }}
\newcommand{\informalconsensuspropertyspace}{\textit{Representation of Preferences }}
\newcommand{\pdmetric}{PD loss}
\newcommand{\pdmetricspace}{PD loss }
\newcommand{\kem}{Kemeny}
\newcommand{\kemspace}{Kemeny }
\newcommand{\weightedKemeny}{Kemeny-Weighted}
\newcommand{\weightedKemenyspace}{Kemeny-Weighted }
\newcommand{\correctperm}{Correct-Pick-A-Perm}
\newcommand{\correctpermspace}{Correct-Pick-A-Perm }
\newcommand{\pickfair}{Pick-Fairest-Perm}
\newcommand{\pickfairspace}{Pick-Fairest-Perm }

\newcommand{\ear}[1]{\textcolor{purple}{EAR: #1}}
\newcommand{\kac}[1]{\textcolor{red}{KAC: #1}}

\title{
MANI-Rank: Multiple Attribute and Intersectional Group Fairness for Consensus Ranking\\ 
\thanks{(*) We thank NSF for support of this research via Grant 2007932.}
}

\author{\IEEEauthorblockN{Kathleen Cachel}
\IEEEauthorblockA{
\textit{Worcester Polytechnic Institute}\\
kcachel@wpi.edu}
\and
\IEEEauthorblockN{Elke Rundensteiner}

\IEEEauthorblockA{
\textit{Worcester Polytechnic Institute}\\
rundenst@wpi.edu}
\and
\IEEEauthorblockN{Lane Harrison}
\IEEEauthorblockA{
\textit{Worcester Polytechnic Institute}\\
ltharrison@wpi.edu}

}

\maketitle

\begin{abstract}

Combining the preferences of many rankers into one single consensus ranking is critical for consequential applications  from hiring and admissions to lending.
\color{black}
While group fairness has been extensively studied for classification, group fairness in rankings and in particular rank aggregation  remains in its infancy.
%underrepresented.
Recent work introduced the concept of fair rank aggregation for combining rankings 
but restricted to the case when candidates have a single binary protected attribute, i.e., they fall into two groups only.
Yet it remains an open problem how to create a consensus ranking that represents the preferences of all rankers while ensuring fair treatment for candidates with multiple protected attributes such as gender, race, and nationality.
In this work, we are the first to define and solve this open Multi-attribute Fair Consensus Ranking (\ourproblemnospace) problem.
\color{black}
As a foundation, we design novel group fairness criteria for rankings, called \fairnesscriterianospace, 
ensuring fair treatment of groups defined by  individual protected attributes and their intersection.
Leveraging the \fairnesscriteria criteria, we develop a series of algorithms that for the first time tackle the \ourproblem problem. 
Our experimental study with a rich variety of consensus scenarios demonstrates our \ourproblem methodology is the only approach to achieve both intersectional and protected attribute fairness while also representing the preferences expressed through many base rankings.
Our real world case study on merit scholarships illustrates the effectiveness of our 
MFCR
%fair consensus ranking
methods to mitigate bias across multiple protected attributes and their intersections. This is an extended version of “MANI-Rank: Multiple Attribute and Intersectional Group Fairness for Consensus Ranking”, to appear in ICDE 2022 \cite{cachel2022manirank}.

\end{abstract}


\section{Introduction}\label{intro}

Rankings are used to determine who gets hired for a job \cite{harwell2019face}, let go from a company \cite{green_2017},  admitted to school \cite{pangburn2019schools}, or rejected for a loan \cite{townson_2020}.
These consequential rankings are often determined through the combination of multiple preferences (rankings)
provided by decision makers into a single representative consensus ranking that best reflects their collective preferences. %The task of combining multiple rankings into a consensus ranking is referred to as \textit{rank aggregation}.
 
However, human and algorithmic decision-makers may generate biased rankings \cite{greenwald1995implicit, o2016weapons, noble2018algorithms}.
Such unfair rankings when combined may escalate into a particularly unfair consensus ranking.
A prevalent type of fairness, called {\it group fairness}, is concerned with the fair treatment of candidates regardless of their 
values for a protected attribute
Protected attributes may be traits such as $Gender$, $Race$, or $Disability$, that are legally protected from discrimination. 
%[CERTIFYING REMOVAL]
More broadly,  protected attributes correspond to  any categorical sensitive attributes for which bias mitigation is desired. 
The problem of incorporating multiple fairness objectives into the process of producing a consensus ranking, namely, \textit{multi-attribute fair consensus ranking} (MFCR), remains open.

\noindent\textbf{Admissions Example.}
Consider an admissions committee ranking applicant candidates for scholarship merit awards as seen
in Figure 1.
First, each of the four committee members, potentially assisted with an AI-screening tool \cite{waters2014grade}, ranks the prospective candidates $-$ illustrated with rankings $r_1$ to $r_4$.
The committee seeks a {\it consensus ranking} for final decision making by consolidating the individual rankings into one single ranking.
For the committee members to accept the outcome, the consensus ranking needs to reflect all
the individual rankings in as much as possible.
% 
% preferences
% of all committee members. %as expressed by their rankings. 
\begin{figure}[t]
\centering{\includegraphics[width=\columnwidth, height=\textheight, keepaspectratio]{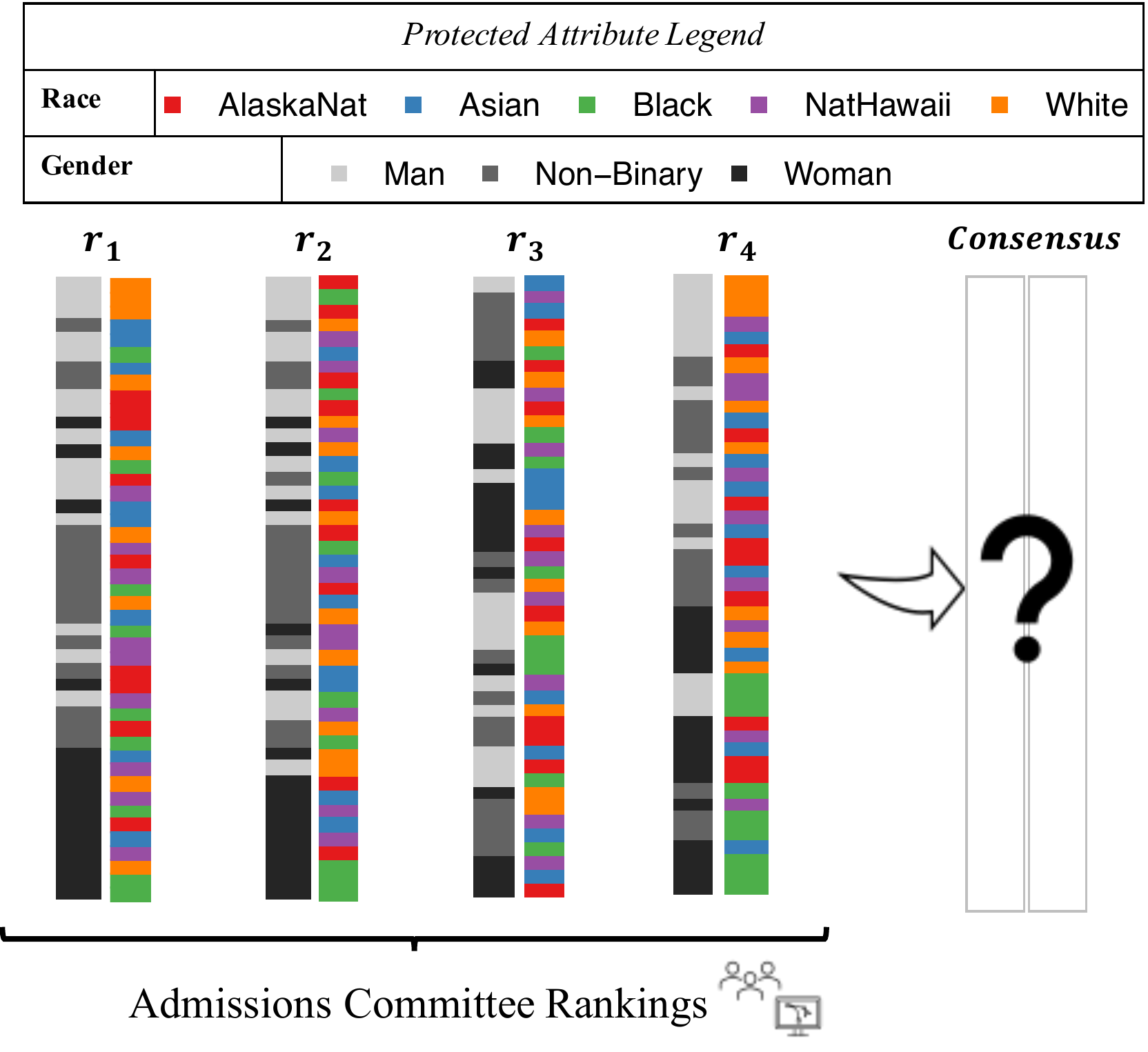}}
\caption{Admissions committee example: $4$ base ranking by four alternate committee members over $45$ candidates with protected attributes $Gender$ and $Race$ to be aggregated into a 'fair' consensus ranking.}
\label{admissions_ex}
\end{figure}

\color{black}
The rankings may conflict substantially $-$ as we can see comparing $r4$ to $r3$ in Fig. 1.
Ranking $r4$ exhibits significant gender and racial bias --- with black candidates and women candidates ranked at the bottom. 
In contrast, $r3$ has relatively even gender and race distributions. 
In fact, $r3$ appears to be the only base ranking that does not have a significant preference for men candidates.
For their decision making, the committee must  ensure their resulting consensus ranking is unbiased with respect to the applicants' Race and Gender.

For this, the committee must define what constitutes \textit{fair treatment} in this setting. This is a  challenge, as group fairness \cite{dwork2012fairness} in rankings with multiple protected attributes is largely unexplored.
If the committee were to only consider Race and Gender independently, will this also mitigate intersectional bias?
Intersectional bias \cite{crenshaw1990mapping}, introduced by legal scholar 
Kimberle Crenshaw, refers to how identities compound to create structures of oppression, and is defined as the combination of multiple protected attributes.
Attempting to treat each protected attribute individually can provide the appearance of fair treatment (e.g., $Gender$ and $Race$ independently), but risks neglect of mitigating intersectional group bias.
For example, black women (frequently at the bottom of $r1, r2$ and $r4$) may not receive fair treatment in the consensus ranking. 
Complicating matters further, individual protected attributes are often independently protected from discrimination via labor laws \cite{ellis2012eu} and civil rights legislation \cite{smith1950discrimination, berg1964equal, green_2017}. 
Thus, if the committee were to
achieve intersectional fairness, it is unclear  if this would also provide
the necessary fair treatment of $Gender$ and $Race$ groups?

Most importantly, the committee is obligated to certify that their final ranking is bias free,
that may require a consensus ranking that is fairer than any of the individual rankings.
The above challenges demand a computational strategy capable of supporting the committee in achieving the fair treatment across multiple protected attributes while also reflecting committee preferences in the fair consensus.

\color{black}
\noindent\textbf{State-of-the-Art and Its Limitations.}
Recent work addressing {\it group fairness with multiple  protected attributes}  focused exclusively on classification and prediction tasks \cite{kearns2018preventing, foulds2020intersectional, hebert2018multicalibration}.
\color{black}
Prior work on group fairness for (single) rankings
%tasks 
has
%exclusively 
addressed only one {\it single facet of fairness}, either  only one protected attribute \cite{yang2017measuring, kuhlman2019fare, beutel2019fairness, geyik2019fairness, zehlike2017fa}
or only the intersection of attributes \cite{yang2020causal}.
Further, in most cases only the restricted case was studied, where this single protected attribute is limited to be binary -- i.e., only two groups \cite{singh2018fairness, yang2017measuring, beutel2019fairness, kuhlman2020rank, kuhlman2019fare}.

Numerous algorithmic strategies exist for {\it combining  base rankings} into a good consensus ranking  \cite{kemeny1959mathematics, ali2012experiments, conitzer2006improved, dwork2001rank, mclean1990borda} -- a task known to be NP-hard in general
\cite{ali2012experiments, dwork2001rank}. 
However, all but one {\it do not consider fairness}. The exception is the recent work by Kuhlman and Rundensteiner  \cite{kuhlman2020rank} on fair rank aggregation, which is limited to providing fair consensus rankings only
for the restricted case when there is one binary protected attribute.
With this restriction, $Race$ would have to be encoded as a binary value,  such as \{white,  non-white\} as opposed to any number of racial categories; while all other protected attributes such as Gender would have to be disregarded.  
%ignored.

\noindent\textbf{Our Proposed Approach.}
In this work, we formulate and then study the problem of multi-attribute fair consensus ranking, namely MFCR, in which we aim to create a fair consensus ranking from a set of base rankings over candidates defined by multiple and multi-valued protected attributes. 
The \ourproblem problem seeks to satisfy dual criteria -- (1.) that all protected attributes and their intersection satisfy a desired level of group fairness and (2.) that the consensus ranking represents the preferences of rankers as expressed by the base rankings. 
We formulate the preference criteria of the MFCR problem through a new  measure called Pairwise Disagreement loss, which allows us to quantify the preferences of rankers not represented in the consensus ranking.
We operationalize the {\it group fairness criteria} of the MFCR problem through interpretable novel fairness criteria we propose called 
Multiple Attribute and Intersectional Rank group fairness (or short,
\textit{MANI-RANK}).
%encompassing both protected attribute fairness and intersectional fairness.

Our formulation of \fairnesscriteria fairness
corresponds to an {\it interpretable unified} notion of fairness for both the protected attributes and their intersection.
Further, this innovation  empowers our family of proposed \ourproblem algorithms  to \textit{tune the degree of fairness} in the consensus ranking via a single parameter. %
% $\Delta$ which we introduce.
%
Thus, our MFCR algorithms achieve the desired level of fairness even among base rankings that may be deeply unfair.
% We design a family of \ourproblem algorithms, starting with an
More precisely, our optimal \ourproblem algorithm called \fairKemenyspace 
% based on linear integer optimization 
elegantly leverages our proposed formulation of \fairnesscriteria as constraints on the exact Kemeny consensus ranking \cite{kemeny1959mathematics}.
We further design a series of polynomial-time \ourproblem solutions based on the efficient consensus generation methods Copeland \cite{copeland1951reasonable}, Schulze \cite{schulze2018schulze}, and Borda \cite{borda1784memoire}.

We conduct a comprehensive experimental study comparing our proposed \ourproblem solutions against a rich variety of alternate consensus ranking strategies from the literature (both fairness-unaware and those that we make  fairness-aware). We  demonstrate that
our solutions are consistently superior for various agreement and fairness conditions. Our experiments also demonstrate the scalability of our proposed methods for large consensus generation problems.
Lastly, we  showcase our \ourproblem solutions in removing bias in a real-world case study involving student rankings for merit scholarships \cite{kimmons}.

\noindent{\bf Contributions.}
Our contributions include the following:
\begin{itemize}
    \item
    We formulate  open multi-attribute fair consensus ranking (MFCR), unifying the competing objectives of bias mitigation and preference representation by adopting a {\it unified  pairwise candidate disagreement model} for both.
    \item
    We design the \fairnesscriteria group fairness criteria, that for the first time {\it interpretably captures} both protected attribute and intersectional group fairness for rankings over candidates with multiple protected attributes.
    \item
    We develop a series of algorithms from 
    MFCR-optimal \fairKemenyspace to polynomial-time \fairCopeland, \fairSchulze, and \fairBordaspace for efficiently solving the open problem of multi-attribute fair consensus ranking. 
    \item

    Our extensive experimental study demonstrates both the efficacy and efficiency of our algorithms, along with the ability to produce real-world fair consensus rankings. We illustrate that only our proposed methodology achieves both group fairness and preference representation over a vast spectrum of consensus problems.
    %\item
    %We illustrate the value of our framework in achieving fair treatment of consensus ranked candidates in two real-world case studies.
\end{itemize}
 
\color{black}
\section{Multi-attribute Fair Consensus Ranking Problem}

\subsection{Preliminaries}\label{prelim}

%\section{Preliminaries and Problem Description}\label{prelim}
\smallskip

\noindent\textbf{Protected Attributes.}
In our problem setting, we are given a database $X$ of $n$ candidates, $x_i \in X $. We assume that each candidate is described by attributes including several categorical {\it protected attributes},
such as, gender, race, nationality, or age.
The set of protected attributes is denoted by $\mathcal{P} = \{p^1,..., p^q \}$, with $q$ protected attributes.
Each protected attribute, say $p^k$, draws values from a domain
of  values  dom($p^k$) =  $\{v_{1}^k,..., v_{q}^k\}$,
with the domain size denoted by $\vert dom(p^k) \vert$, or, in
short, $\vert p^k \vert$.
For instance, the domain of the $k$-th protected attribute $Gender$
is composed of the three values $Man, Woman, Non-binary$.

The protected attributes $\mathcal{P}$ combined in a Cartesian product $Inter = {p^1 \times ...\times p^q}$ forms what we call an \textit{intersection} \cite{crenshaw1990mapping}. 
The cardinality of the intersection, i.e., the number of all its possible 
value combinations, is $\vert Inter \vert = \vert p^1 \vert *... * \vert p^q \vert$.
We denote the value of candidate $x_i$'s $k$-th protected attribute by $p^k(x_i)$ and their intersection value as $Inter(x_i)$.

For each value $v_{j}^k$ of a protected attribute $p^k$, there is a {\it protected attribute group} composed of all candidates $x_i \in X$  who all have the same value $v_{j}^k$ for the protected attribute $p^k$.

\smallskip
\noindent\textbf{Definition 1} (Protected Attribute Group) \textit{Given a candidate database $X$ and a value $v_j^k$ for a protected attribute $p^k$,  the \textbf{protected attribute group} for value $v_j^k$ is:}
$$
 G_{(p^k:v_j^k)} = \{ x_i \in X:  p^k(x_i)  = v_j^k. \}\label{protected attribute group}
$$
%For brevity, we  refer to the group of candidates in $X$ who all have value $v_{j}^k$ for the protected attribute $p^k$ simply by $G_{(k:j)}$ instead of $G_{k:v_j^k}$.
For brevity, we  refer to the protected attribute group $G_{p^k:v_j^k}$ by   $G_{(k:j)}$, when possible without ambiguity.
For instance, $G_{(Gender:Woman)}$ denotes the group of all women in $X$.
As candidates in $X$ are defined by their intersectional identity, an \textit{intersectional group} then corresponds to all candidates $x_i \in X$ who share the same values across all protected attributes.

\smallskip
\noindent\textbf{Definition 2} (Intersectional Group) \textit{Given a candidate database $X$ and values ${v_j^1, v_j^2, ..., v_j^q}$ for  protected attributes set $\mathcal{P}$, the \textbf{intersectional group} for values ${v_j^1, v_j^2, ..., v_j^q}$  is:}
$$
\begin{gathered}
 InterG_{(p^1:v_j^1),..., (p^q:v_j^q)} =\\ \{ x_i \in X:  (p^1(x_i)  = v_j^1) 
  AND   ...  AND  (p^q(x_i)  = v_j^q) \}.
\label{inter group}
\end{gathered}
$$
For brevity, a group of candidates in $X$ sharing an intersection value, the $j$-th intersection value, is denoted $InterG_{j}$.

\noindent\textbf{Notion of Group Fairness.}
In this work, we aim to achieve the group fairness notion of \textit{statistical parity} \cite{dwork2012fairness}.
First proposed in binary classification  \cite{dwork2012fairness}, and more recently the focus of fair learning-to-rank methods for a binary protected attribute \cite{kuhlman2019fare, singh2018fairness, yang2017measuring, zehlike2017fa}, {\it statistical parity} is a requirement stipulating candidates must receive an equal proportion of the positive outcome regardless of their group membership in a protected attribute.

\smallskip
\noindent\textbf{Definition 3} (Statistical Parity) \textit{Given a dataset $X$ of candidates sharing different values of protected attribute $p$, and a binary classifier $\hat{Y}$  with positive outcome $\hat{Y} = 1$. The predictions of $\hat{Y}$ are fair with respect to $p$ if candidates with different values of $p$ have a $Prob(\hat{Y} = 1)$ directly proportional to their share of $X$.}

\noindent\textbf{Base Rankings Model User Preferences.}
 Our problem contains $m$ rankers $-$ where rankers (human, algorithmic, or some combination thereof) express preferences over the  database $X$ of candidates. Each ranker's preferences over $X$ are expressed via a ranking. A {\it ranking} is a strictly ordered permutation $ \pi = [x_1 \prec x_2 \prec x_i \prec ... \prec x_n]$ over all  candidates in $X$.
Here $x_i \prec_\pi x_j $ implies candidate $x_i$ appears higher in the ranking $\pi$ than $x_j$,  where $1$ is the top or best ordinal rank position and $n$ the least desirable. The collection of all possible rankings over the database of $X$ of $n$ candidates, denoted $S_n$, corresponds to the set of all possible permutations of $n$ candidates. The $m$ rankings produced by \textit{the m rankers} creates a set $R \subseteq S_n$, which we refer to as  \textit{base rankings}.

\noindent\textbf{Consensus Ranking.}
From the base rankings $R$, we wish to generate a ranking that represents the preferences of the rankers, namely, a \textit{consensus ranking}.
A \textit{consensus ranking} $\pi^C$  is the ranking closest to the set of base rankings $R$, such
that a distance function $dist$ is minimized. Finding a consensus ranking corresponds
to traditional rank aggregation task \cite{brandt2012computational}.

\smallskip
\noindent\textbf{Definition 4} (Consensus Ranking)
 \textit{
Thus, a \textbf{consensus ranking} $\pi^C$ from a set of base rankings $r_i \in R$ is defined by:
\begin{equation}
    {\pi^C} = \underset{\pi \in S_n}{\mathrm{argmin}} \frac{1}{\vert R\vert} \sum_{r_i \in R} dist(\pi, r_i).
\label{rankaggregation}
\end{equation}
}

\noindent\textbf{Multi-attribute Fair Consensus Ranking (\ourproblemnospace)}.
Our problem of creating a fair consensus ranking from many rankers' preferences over candidates defined by multiple and multi-valued protected attributes has \textit{two components}.
The fairness component aims to treat \textit{all candidates equally regardless of group membership} in protected groups (Definition 1) or intersectional groups (Definition 2).
In order for the fairness criteria of our problem to be meaningful 
%and useful 
in practice (i.e, in the form of targets such as the "80\%" rule championed by US Equal Employment Opportunity Commission (EEOC) \cite{us1979questions}), we define "fair" as an \textit{application-specified desired level of fairness} in the consensus ranking.

In Section \ref{fairnesscriteriasect}, we propose novel group fairness criteria which we integrate into the MFCR problem (Definition 11) as a concrete target (i.e, \textit{the application can select a proximity to perfect statistical parity \cite{dwork2001rank}}). Thus, the MFCR problem 
%\textit{
encompasses the creation of consensus rankings that may need to be fairer than the fairest base ranking.
%}.

The preference representation component   
of our problem ensures that all rankers see their preferences reflected in the fair consensus ranking in as much as is possible, and thus they can be expected to  accept the consensus ranking.
We note that even inside biased base rankings critical preference information is encoded, such as the orderings of candidates within the same group. Thus in Section \ref{prefcriteriasection}, we propose a new measure to quantify the preferences of the base rankings $R$ that are not represented in the fair consensus ranking, called \pdmetric. We integrate this measure into the MFCR problem (Definition 11) as criteria to be minimized in the generation of a fair consensus ranking. Doing so ensures that a fair consensus ranking does not prioritize certain rankers above others.

\subsection{Proposed Group Fairness Criteria for Multi-attribute Fair Consensus Ranking (MFCR)}\label{fairnesscriteriasect}

\noindent\textbf{Proposed Group Positive Outcome Metric.}
%\color{black}
%\section{Proposed Group-based Positive Outcome For Multi-Attribute Fair Rank Aggregation}
To operationalize the group fairness component of our problem (Section \ref{prelim}), we design a measure for capturing how fairly a group is being ranked.
Our insight here is to define \textit{fair treatment} of a group via a constant value, thus making the interpretation across group sizes comparable. 

This allows us to define statistical parity (Definition 3)
based on all groups having this same value, thus formulating group fairness for multiple and multi-valued protected attributes into a single easy-to-understand measure.

%\subsection{Proposed Rank Aggregation Positive Outcome Metric}
\color{black}
For many applications, the entire ranking matters $-$ bottom ranked candidates may lose out on consequential outcomes such as funding, resource and labor divisions, or decreased scholarships, if they were placed even somewhat lower.
%For this, we leverage statistical parity as defined in Definition 3, based on leveraging
To capture how a group is treated though a ranking we utilize {\it pairwise comparisons of candidates}.
Intuitively, the more pairs a candidate is favored in, the higher up in the ranking the candidate appears. 
Any ranking  $\pi$ over $n$ candidates can be decomposed into pairs of candidates ($x_i, x_j$) where $x_i \prec_\pi x_j$.
The total number of pairs in a ranking over a database $X$ of $n$ candidates is:
\begin{equation}
    \omega(X) = (n(n - 1))/{2}.\label{totalpaircount}
\end{equation}
{\it As statistical parity requires groups receive an equal proportion of the positive outcome \cite{dwork2012fairness}}, we cast positive outcome as being favored in a mixed pair $-$ where mixed refers to candidates in a pair associated with two distinct protected groups according to protected or intersection attributes.
For instance, a pair comparing two woman candidates is not a mixed pair, while a pair with a man and woman is a mixed pair.
We denote the number of mixed pairs for a group $G$ (where $G$ is  a protected attribute group $G_{k:j}$ or  intersectional group $InterG_j$) in a ranking over $\vert X \vert$ candidates with $\vert G \vert \leq \vert X \vert$ 
as:
\begin{equation}
    \omega_{M}(G, \pi) = \vert G \vert (\vert X \vert - \vert G \vert).
    \label{mixedpairs}
\end{equation}
The total number of mixed pairs for a protected attribute $p^k$ or intersection $Inter$ in a ranking over $\vert X \vert$ candidates is:
\begin{equation}
    \omega_M(X, \pi) = \omega(X, \pi) - \sum_{G_{*:i} \in X}\omega(G_{*:i}, \pi),
    \label{totalmixedpairs}
\end{equation}
where $* = k$ or $* = inter$ for the respective attribute.

% Beyond groups defined by a binary valued attribute, 

We  design {\it our measure of positive outcome allocation,} called a group's  Favored Pair Representation ($FPR$) score.

\smallskip
\noindent\textbf{Definition 4} (Favored Pair Representation) \textit{The \textbf{FPR} for a ranking $\pi$ over candidate database $X$ for a group of candidates $G$ $\subset  X$, where $G$ is either $G_{k:j}$ or $InterG_j$, is:}
$$
 FPR_{G}(\pi) = \sum_{x_i \in G} \sum_{x_l \notin G}\frac{countpairs(x_i \prec x_l)}{\omega_{M}(G, \pi)}.\label{FPR}
$$

One critical property of this 
 $FPR$ score is that it is easy to explain and interpret. It ranges from $0$ to $1$.
When $FPR = 0$, the group is entirely at the bottom of the ranking.
When  $FPR = 1$, the group is entirely at the top of the ranking.
By design,  when $FPR = 1/2$, the group receives fair treatment in the ranking, i.e., a directly proportional share of favored rank positions. 

Next, we assure that our $FPR$ measure handles
{\it  groups defined by 
multi-valued attributes.}
For that,
we put forth that when the attribute has multiple values, we must divide by the number of mixed pairs containing that specified group (Equation \eqref{mixedpairs}) as opposed to  the total number of mixed pairs in the ranking
(Equation \eqref{totalmixedpairs}).
Our design 
unlike prior work \cite{kuhlman2019fare, kuhlman2020rank},guarantees the critical property of the $FPR$ measure that $1/2$ is a fair positive outcome allocation even for {\it multi-valued attributes}  groups.

We observe  that the $FPR$ measure allows us to compare the fair treatments of {\it groups of different sizes} purely based on their $FPR$ scores. 
Better yet, when all groups receive a proportional share of the positive outcome (i.e, $FPR = .5$) then all groups receive an equal proportion of the positive outcome, thus satisfying statistical parity. 
Thus, the $FPR$ metric elegantly allows us to check for perfect statistical parity simply by having an $FPR$ score of $0.5$ for all groups.

\color{black}

\noindent\textbf{Proposed Unified Multi-Attribute Group Fairness Criteria.}
%\subsection{Unified Multi-Attribute Group Fairness Criteria}
We now propose our formal definition of the group fairness criteria of our MFCR problem. %(Section \ref{prelim}). 
%%%%
\color{black}
Our key design choice is to specify group fairness at the granularity of the attribute $-$ as opposed to fairness at the group level.
In this way, we provide the ability to \textit{tune the degree of fairness} in every protected attribute and in the intersection.
Intentionally, we do not design criteria per group.
As in the multiple protected attribute setting, even a \textit{handful of attributes and their intersection} can create a \textit{huge number of groups} that otherwise  would have to be individually parameterized and interpreted.

For the group fairness property of MFCR, we introduce a new measure to quantify statistical parity for protected attributes $p^k$ as described below.

\smallskip
\noindent\textbf{Definition 5} (Attribute Rank Parity) \textit{The \textbf{Attribute Rank Parity (ARP)} measure for the $k$-th protected attribute $p^k$ in ranking $\pi$ over candidate database $X$ is:}
$$
ARP_{p^k}(\pi) = \underset{\forall ~(G_{k:i}, G_{k:j}) \in X}{\mathrm{argmax}} \vert FPR_{G_{k:i}}(\pi) - FPR_{G_{k:j}}(\pi) \vert
\label{arp}
$$
This ARP measure simplifies the task of tuning the degree of fairness for a protected attribute.
Namely, when the $ARP_{p^k} = 1$, then the protected attribute is maximally far from statistical parity. Meaning, one group corresponding to a value in the dom($p^k$) is entirely at the top of the ranking, while a second group is entirely at the bottom of the ranking.
When $ARP_{p^k} = 0$, perfect statistical parity is achieved for attribute $p^k$.

Similarly, we now formulate intersectional fairness, 
which corresponds to criteria $ii$ of our 
% fair consensus generation 
problem.
% via closeness to statistical parity below.

% We refer to this measure as {\it Intersectional Rank Parity} ($IRP$).

\smallskip
\noindent\textbf{Definition 6} (Intersectional Rank Parity) \textit{The \textbf{Intersection Rank Parity (IRP)} measure for the intersection $Inter$ determined from the set of protected attributes $\mathcal{P}$ in ranking $\pi$ over candidate database $X$ is:}

$$
\begin{gathered}
    IRP(\pi) = \\ \underset{\forall (InterG_{i}, InterG_{j}) ~\in ~X}{\mathrm{argmax}} \vert FPR_{InterG_{i}}(\pi) - FPR_{InterG_{j}}(\pi) \vert\label{irp}
\end{gathered}
$$

We now have {\it two easy-to-use and interpretable measures}  $ARP$ and $IRP$ that directly map to the fair treatment of protected attribute groups (Definition 1) and intersectional groups (Definition 2) in the MFCR problem.
We unify these objectives into one fairness notion, which we call {\it Multiple Attribute and Intersectional Rank} (\fairnesscriterianospace) group fairness. \fairnesscriteria is applicable to consensus and to single rankings alike over candidate databases with multiple protected attributes.
%This new measure is applicable to any ranking - consensus or regular.
% non-consensus alike.
\color{black}

We introduce the threshold parameter $\Delta$ for \textit{fine-tuning a desired degree of fairness}.
$\Delta$ 
represents the {\it  desired (or required) proximity to statistical parity}. Equation \eqref{Arp_multifair} below models this for every protected attribute and Equation \eqref{Irp_multifair}  for the intersection.
This is carefully designed to be easy to interpret, namely,   {\it  when $\Delta$ is close to $0$, the ranking approaches statistical parity for all protected attributes and intersection. }

\noindent\textbf{Definition 7} (Multiple Attribute and Intersection Rank $-$ MANI-Rank) \textit{\textbf{MANI-Rank Group Fairness} 
% criteria
for rankings of candidates with multiple protected attributes is defined as:}
\begin{equation}
ARP_{p^k}(\pi) \leq \Delta~(\forall~p^k \in \mathcal{P})\label{Arp_multifair}
\end{equation}
\begin{equation}
~IRP(\pi) \leq \Delta\label{Irp_multifair}
\end{equation}

Per design, the \fairnesscriteria fairness criteria  result in perfect protected attribute and intersectional statistical parity,
when $\Delta = 0$.
This follows  from  definitions of $ARP_{p^k}$ and $IRP$.

\noindent\textbf{Customizing Group Fairness.} 
In most real-world settings, equal degrees of fairness for protected attributes and their intersection is desirable, which we model by 
choosing the same degree of fairness threshold $\Delta$ in Definition 7.
However, our \fairnesscriteria criteria allows for
{\it 
applications to set up different thresholds} tailored to
each protected attribute ($\Delta_{p^k}$) or the intersection ($\Delta_{Inter}$).
Additionally, Definition 7 can be modified to \textit{handle alternate notions of intersection} by adjusting the intersectional groups (Definition 2) to be a desired subset of protected attributes (as opposed to the combination of all protected attributes).
Likewise, Definition 7 can be extended to \textit{support specific subsets of protected attribute combinations} by adding to Definition $7$  an additional equation for every desired subset of protected attributes, such as $IRP_{subsets of \mathcal{P}}(\pi) \leq \Delta$. We note, as evidenced by our empirical study in Section \ref{constraint_analysis}, in order for a protected attribute or intersection to be guaranteed protected at a desired level of fairness it must be constrained explicitly.

\subsection{Proposed Representation Criteria for Multi-Attribute Fair Rank Aggregation}\label{prefcriteriasection}
We also propose to model the degree to which the consensus ranking captures the preferences of all the rankers corresponding to the preference representation component of MFCR.
For this, we measure the distance between the base rankings $R$ and the fair consensus ranking $\pi^{C*}$. 
We intentionally select pairwise disagreements as a distance measure because this allows us to \textit{interpretably measure} how many preferences, i.e., candidates comparisons, are not met in the fair consensus ranking.  
We propose to do this by summing the \textit{Kendall Tau} \cite{kendall1938new} distances between $\pi^{C*}$ and every base ranking. The Kendall Tau distance in Definition 8 is a distance measure between rankings. It counts the number of pairs in one ranking that would need to be inverted (flipped) to create the second ranking, thereby counting pairwise disagreements between two rankings.
\smallskip

\noindent\textbf{Definition 8} (Kendall Tau) \textit {Given two rankings $\pi_1, \pi_2 \in S_n$, the {\bf Kendall Tau distance}
%measure 
between them is:}
$$
\begin{gathered}
 dist_{KT}(\pi_1,\pi_2) = \\
 \vert \{\{x_i, x_j\}\in X: x_i \prec_{\pi_1} x_j \text{ and } x_j \prec_{\pi_2} x_i  \}\vert\label{kt}.
\end{gathered}
$$
Then we normalize the pairwise disagreement count by the total number of candidate pairs represented in the base rankings $R$. This leads us to Definition 9 of our proposed preference representation metric {\it Pairwise Disagreement loss} (\pdmetric). 
\color{black}
\smallskip
\noindent\textbf{Definition 9} (Pairwise Disagreement Loss) \textit{Given a set of base rankings $R$ and a consensus ranking $\pi^{C}$, the \textbf{pairwise disagreement loss (\pdmetric)} between $\pi^{c}$ and $R$ is:}
$$
\begin{gathered}
 PD\ Loss(R,\pi^{C}) = \frac{\sum_{r_i \in R} dist_{KT}(\pi^{C}, r_i)}{\omega(X, \pi^C)*\vert R\vert}
\end{gathered}
$$
We now have a measure that expresses  how many preferences of  rankers are not captured in a fair consensus ranking. By design, \pdmetricspace is between the values of $0$ and $1$, with $0$ offering the interpretation that \textit{every pairwise preference} in the base rankings is represented in the fair consensus ranking (i.e., all the base rankings are the same and match exactly the fair consensus ranking) and \pdmetric \; is $1$ when no single pairwise preference in the base rankings is  represented in the fair consensus ranking.

\subsection{\ourproblem Problem: Multi-Attribute Fair Consensus Ranking}\label{formalprobsect}
Pulling together our proposed group fairness and preference representation models, we now are ready to formally  characterize   our fair consensus ranking problem.
% when the candidate database contains multiple and multi-valued protected attributes. 

\smallskip
\noindent\textbf{Definition 10} (Multi-Attribute Fair Consensus Ranking - \ourproblem) \textit{Given a database of candidates $X$ defined by multiple and multi-valued protected attributes $\mathcal{P}$, a set of base rankings $R$, and proximity to statistical parity parameter $\Delta$,
the \textbf{ multi-attribute fair consensus ranking (MFCR)} problem is to create a consensus ranking $\pi^{C*}$, that meets two criteria:}

\noindent{\bf \mfracritfair}:
\begin{itemize}
    \item satisfies {\bf \fairnesscriteria group fairness} (Definition $7$) subject to parameter $\Delta$, and,
\end{itemize}
{\bf \mfracritbaser:}
\begin{itemize}
    \item minimizes \pdmetricspace (Definition $9$) between $R$ and $\pi^{C*}$.
\end{itemize}
Our problem formulation emphasizes the dual ability to specify a desired level of group fairness in the consensus ranking while minimizing unrepresented ranker preferences in the consensus ranking. We note that, by design, this allows for applications to create a consensus ranking \textit{fairer than all the base rankings} by setting a low $\Delta$ parameter.

\color{black}

\begin{figure}[t]
\centering{\includegraphics[width=.99\columnwidth, keepaspectratio]{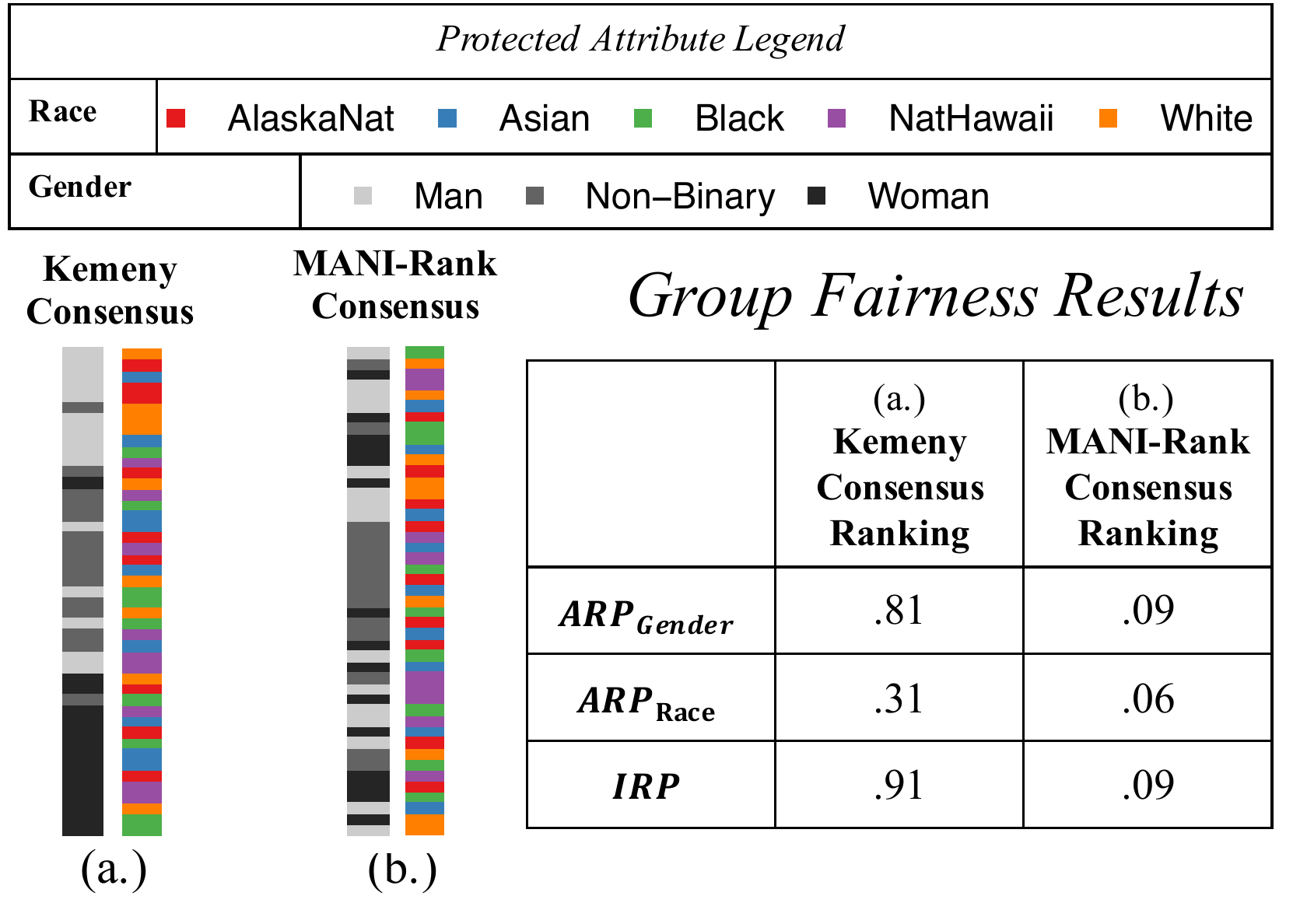}}
\caption{Admissions committee example: (a) Kemeny consensus ranking and (b) \fairnesscriteria consensus ranking}
\label{admissions_ex_ra}
\end{figure}

% \subsection{MFRA Applied to Motivation Admissions Example}
\noindent{\bf \ourproblem Applied to Motivation Admissions Example.} Returning to the task faced by the Admissions Committee from Section \ref{intro}, we demonstrate how the committee can apply the MFCR problem to create a fair consensus ranking of applicant candidates.
Figure \ref{admissions_ex_ra} illustrates the consensus ranking generated for admissions decisions with and without \fairnesscriteria group fairness.
The ranking \ref{admissions_ex_ra}a is determined from premininent consensus ranking method, Kemeny \cite{kemeny1959mathematics}.
We see that it exhibits significant bias with respect to Gender as men are clustered at the top.
Likewise, the intersectional bias is substantial due to  white men being significantly advantaged in the ranking.
In contrast, ranking \ref{admissions_ex_ra}b, generated with \fairnesscriteria $\Delta = .1$, has $ARP$ and $IRP$ scores nearly at perfect statistical parity - indicating the promise of our proposed formulation to remove Gender, Race ~and intersectional bias. 
Apply the \ourproblem framework helps the admissions committee to create a fair consensus ranking; as otherwise biases present in  base rankings would be reflected in the final applicant ranking.

\section{Algorithms For Solving Multi-attribute Fair Consensus Ranking}\label{algosection}

We propose a family of algorithms for solving the \ourproblem problem. 
For specific algorithms, we utilize an 
a \textit{precedence matrix}  representation $W$  of the base rankings $R$
that captures all pairwise relationships in  $R$.
Put differently, $W$'s entries represent \textit{pairwise disagreements}.

\color{black}
\noindent\textbf{Definition 11} (Precedence Matrix) \textit{Given a database $X = \{x_1, x_2, ..., x_n\}$ of candidates, and set of base rankings $ R$,  the \textbf{precedence matrix} $W = [W_{x_a,x_b}]_{a,b = 1,...,n}$ is defined by}: 
$$
 {W_{x_a,x_b}} = \sum_{r_i \in R} \mathbbm{1}{(x_b \prec_{r_i} x_a)}
 \label{precedence}
$$
\textit{where $\mathbbm{1}$ is the indicator function, namely, equal to $1$ when $x_b \prec_{r_i} x_a$, $0$ otherwise.}

\color{black}
\subsection{\fairKemenyspace Strategy for Solving Our \ourproblem Problem}\label{exact}
We design a method called \fairKemenyspace which optimally satisfies all criteria of our \ourproblem problem.

\textit{Kemeny} corresponds to a specific instantiation of finding a consensus ranking, where the pairwise disagreement metric, the Kendall-Tau distance, is minimized between the consensus $\pi^C$ and base rankings $R$.
Kemeny is a Condorcet  \cite{fishburn1977condorcet} method.
Consensus ranking methods, such as Kemeny \cite{kemeny1959mathematics}, that satisfy the Condorcet criteria, order candidates by how preferred (using head-to-head pairwise candidate comparisons) there are amongst rankers.
%\ear{Was calling it voters here ok, or confusing?}
% voters.
Thus Condorcet methods naturally align with our \mfracritbaser. 
%Thus, it is a natural choice for satisfying the \mfracritbaser.
By incorporating \fairnesscriteria as a set of constraints, we can  leverage the exact Kemeny formulation $-$ proven to return the Kemeny optimal consensus ranking \cite{ali2012experiments, conitzer2006improved,schalekamp2009rank}.
\fairKemenyspace models \fairnesscriteria group fairness as constraints in the exact Kemeny Integer Program to satisfying \mfracritfairspace optimally. 
\color{black}

\begin{algorithm}

\renewcommand{\algorithm}{}
\LinesNotNumbered
\begin{equation}
    \textrm{Minimize} \sum_{\forall (x_a, x_b) \in X}Y_{(x_a, x_b)}W_{(x_a, x_b)}\label{objective}
\end{equation}
\begin{equation}
    \textrm{ subject to:}~ Y_{(x_a, x_b)} \in \{0,1\}\label{binaryvar}    
\end{equation}
\begin{equation}
    \textrm Y_{x_a,x_b} + Y_{x_b,x_a} = 1, \forall x_a, x_b \label{strictorder}    
\end{equation}
\begin{equation}
    \textrm Y_{x_a,x_b} + Y_{x_b,x_c} + Y_{x_c,x_a} \leq 2, \forall x_a, x_b, x_c \label{transitivity}    
\end{equation}
\begin{equation}
\begin{gathered}
    \vert \sum_{x_a \in G_{k:i}} \sum_{x_b \notin G_{k:i}}(\frac{1}{\omega_m(G_{k:i})}Y_{(x_a,x_b)}) - \\ \sum_{x_c \in G_{k:j}} \sum_{x_d \notin G_{k:j}}(\frac{1}{\omega_m(G_{k:i})}Y_{(x_c,x_d)})\vert \leq \Delta, \\\forall (G_{k:i}, G_{k:j}) \in X, \forall ~p^k \in \mathcal{P}\label{ARP}    
\end{gathered}
\end{equation}
\begin{equation}
\begin{gathered}
    \vert \sum_{x_a \in InterG_{i}} \sum_{x_b \notin InterG_{i}}(\frac{1}{\omega_m(InterG_{i})}Y_{(x_a,x_b)}) - \\ \sum_{x_c \in InterG_{j}} \sum_{x_d \notin InterG_{j}}(\frac{1}{\omega_m(InterG_{i})}Y_{(x_c,x_d)})\vert \leq \Delta, \\\forall (InterG_{i}, InterG_{j}) \in X\label{IRP}    
\end{gathered}
\end{equation}
\caption{\fairKemeny}
\end{algorithm}
\color{black}
In the formulation of \fairKemenyspace above, matrix $Y$ specifies the pairwise order of candidates in the consensus ranking $\pi^{C*}$ and matrix $W$ represents the precedence matrix from Definition 11. The objective function in Equation \eqref{objective} formulates the Kemeny criteria, minimizing the number of pairwise disagreements between  base rankings $R$ and $\pi^{C*}$.

As shown in Conitzer et al. \cite{conitzer2006improved}, the first three constraints, Equations \eqref{binaryvar}, \eqref{strictorder}, \eqref{transitivity} enforce a valid ranking (no cycles, not multiple candidates in the same position, or no invalid pairwise orderings).

Next, our formulation of \fairnesscriteria group fairness
is modeled by the constraints in Equation \eqref{ARP} and Equation \eqref{IRP} 
which enforces group fairness for every protected attribute as well as 
 intersectional group fairness, respectively.
These constraints leverage our  formulation of $ARP$ (Definition 5) and $IRP$ (Definition 6) constraining the maximum absolute difference in $FPR$ scores for all groups representing values in the dom($p^k$) for each protected attribute and in dom($Inter$).

\noindent\textbf{Complexity  of \fairKemenyspace Solution.} General (fairness unaware) Kemeny is an NP-hard problem \cite{kemeny1959mathematics, dwork2001rank}, though tractable in smaller candidate databases. Our \fairKemenyspace  method inherits this complexity.
Our \fairnesscriteria criteria formulated via Equation (11) adds ${\binom{\vert dom(p^k) \vert}{2}}$ and Equation (12) adds ${\binom{\vert dom(Inter) \vert}{2}}$ constraints. Yet, our empirical study in Section \ref{scalestudysect} confirms that in practice the runtime of \fairKemenyspace is not substantially greater than that of the traditional 
%fairness-unaware 
Kemeny, and both can  handle thousands of base rankings.

\subsection{\fairCopeland, \fairSchulze, and \fairBordaspace}
% for Solving \ourproblem}

Aiming to handle larger candidates databases than \fairKemenyspace or Kemeny,
we now design a series of algorithms 
%that solve the \ourproblem problem while offering increasingly scalable solutions.
utilizing polynomial time consensus generation methods, 
They all
% designs of the subsequent algorithms 
utilize a novel pairwise bias mitigation algorithm we propose, called \makefair, specifically designed to efficiently achieve \mfracritfairspace while minimizing increases  \pdmetricspace caused by introducing fairness.

\noindent\textit{\textbf{\makefairspace}} takes as input a consensus ranking $\pi^c$ to be corrected. %It is utilized by the subsequent solutions to ensure the consensus ranking satisfies \mfracritfair.
Initially, it determines the $FPR$ and $IRP$ scores of the protected attributes and intersection, checking if \mfracritfairspace is satisfied with respect to the $\Delta$ semantics.
%parameter.
\color{black}
When this condition is not true, the algorithm 
determines the attribute (either a protected attribute or the intersection) that has the highest $ARP$ or $IRP$ score.
This attribute is now ``the attribute to correct". By honing in on the least fair attribute we aim to minimize pairwise swaps - thus minimizing \pdmetric.
Within the attribute to correct, the group with the highest
and lowest $FPR$ score is said to be $G_{highest}$ and $G_{lowest}$, respectively.

Within the group $G_{highest}$, the candidate lowest in the ranking is selected as $x_{G_{h}}$.
Then the ordered mixed pairs %(ordered based on increasing ordinal rank position of the unfavored candidate)
of $x_{G_h}$ are searched to determine the first unfavored candidate, $x_{G_l}$, in the list of mixed pairs which belong to $G_{lowest}$.
If there is no $x_{G_l}$ candidate then the $x_{G_h}$ candidate is altered to be the first candidate in $G_{highest}$ higher than the original $x_{G_h}$ candidate.
We intentionally choose to redefine the $x_{G_h}$ candidate as opposed to  $x_{G_l}$. The reason is to
 \textit{enforce re-positioning candidates into positions at the top of the ranking}, making fewer, but more impactful swaps. This helps to minimize the increase in \pdmetricspace caused by this fairness mitigation process. 
When the pair $x_{G_h} \prec_{\pi^*} x_{G_l}$ is found, the two candidates are swapped --  resulting in $x_{G_l} \prec_{\pi^C} x_{G_h}$.
Each pair swap is guaranteed to lower the $FPR$ score of $G_{highest}$ and increase the $FPR$ score of $G_{lowest}$, thus increasing proximity to statistical parity for the attribute to correct.
The algorithm terminates once all protected attributes and the intersection are below $\Delta$, and returns the corrected ranking $\pi^C$ as fair consensus ranking $\pi^{C*}$.

\begin{algorithm}
\renewcommand{\algorithm}{}
\color{black}
\DontPrintSemicolon
  \KwInput{consensus ranking $\pi^C$, candidate database $X$, thresholds $\Delta$}
  \KwOutput{fair consensus ranking $\pi^{C*}$}
  \ForAll{$p^k \in \mathcal{P}~\textrm{and}~ inter$}{
    \textrm{determine} $ARP$ scores and $IRP$ scores
  }
  \While{all $ARP$ scores and $IRP$ scores \textgreater $\Delta$}{
    \textrm{atr}~=~$p^k \in \mathcal{P}~\textrm{or}~ Inter$ \textrm{with max} $IRP/ARP$\tcp*{entity to correct}
    $G_{highest}$~=~group of \textrm{atr with max $FPR$ score}\\
    $G_{lowest}$~=~group of \textrm{atr with min $FPR$ score}\\
    $x_{G_h}$ ~=~ \textrm{lowest $x_i \in G_{highest}$}\\
    \eIf{($x_{G_h} \prec x') \in \omega_m(x_{G_h})$ \textrm{ s.t} $x' \textrm{is the highest }  x' \in G_{lowest}$}{
    $x_{G_l}$~=\textrm{ is the highest } $x' \in G_{lowest}$\\
    \tcc{find next highest $x_{G_h}$}
    }{
    $x_{G_h}$ ~=~ \textrm{next lowest} $x_i \in G_{highest}$ \textrm{s.t.} $(x_{i} \prec x') \in \omega_m(x_i)$ s.t. $x'$ is the highest $x' \in G_{lowest}$ \\
    $x_{G_l}$~=\textrm{ is the highest } $x' \in G_{lowest}$
    }
  swap  $x_{G_l}$ and $x_{G_h}$ 
  }
  $\pi^{C*}$ $=$ $\pi^C$\\
  \textbf{return}~$\pi^{C*}$
\caption{\makefair}
\end{algorithm}

\noindent{\bf Complexity of \makefairspace Algorithm.}
\makefairspace  determines the $ARP$ scores for all protected attributes and the $IRP$ score. Each $ARP$ and $IRP$ score is calculated with one traversal of $\pi^{C}$ to determine the $FPR$ scores that $ARP$ and $IRP$ are calculated from. Thus each score computation costs $O(n)$ work. This work is done before each swap. In the worst case, the algorithm would flip $\omega(X) = n*(n-1)/2$ pairs. Thus, assuming $|\mathcal{P}|$ protected attributes and one intersection, the resulting worst case runtime is $O(n^2*(|\mathcal{P}|+1) *n)$. Additionally, the runtime can be reduced by adjusting the $\Delta$ parameter 
as will be empirically illustrated in Section \ref{scalestudysect}.
\\
\textbf{\textit{\fairCopelandspace Solution for MFRA.}}
We create \fairCopelandspace based on Copeland \cite{copeland1951reasonable}, because the later is 
the fastest (polynomial) pairwise consensus rank generation method \cite{schalekamp2009rank}.
Copeland \cite{copeland1951reasonable} creates a consensus ranking that ranks candidates in descending order by the number of pairwise contests a candidate has won (a tie is considered a "win" for each candidate). Intuitively, this can be understood through our precedence matrix $W$, where the number of pairwise contests candidate $x_b$ wins is $\sum_{x_a \in X}\mathbbm{1}( W_{x_a, x_b} \geq W_{x_b, x_a})$. Our \fairCopelandspace method satisfies \mfracritbaserspace by producing the Copeland consensus, ordering candidates by descending number of wins in pairwise contests, then correcting to satisfy \mfracritfairspace with $\Delta$ using \makefair. The algorithm’s pseudocode is provided in the supplement \cite{supplement}.

\noindent\textbf{Complexity  Of \fairCopeland.} \fairCopelandspace takes $O(n^2*|R|)$ to create the precedence matrix $W$, $O(n^2)$ to check $W$ to determine the winners of pairwise contests, $O(nlogn)$ time to sort the candidates by the number of contests won, and in the worst case \makefairspace takes $O(n^2*(|\mathcal{P}|+1) *n)$ to fairly reorder candidates. 

\noindent\textbf{\textit{\fairSchulzespace Solution for MFCR.}} 
We introduce \fairSchulzespace as it is polynomial-time, Condorecet, and pairwise like \fairCopeland, and in addition its fairness-unaware method is extremely popular in practice.
Schulze is used to determine multi-winner elections in over 100 organizations worldwide (such as Wikimedia Foundation to elect a board of Trustees, Political Parties, Gentoo, Ubuntu, and the Debian Foundation, see \cite{schulze2018schulze} for a full list). The Schulze \cite{schulze2018schulze} method generates a consensus ranking via pairwise comparisons which naturally helps address \mfracritbaser.

The Shulze rank aggregation method first determines the precedence matrix $W$. Then $W$ is treated as a directed graph with weights representing the pairwise agreement counts between every pair of candidates. Next, the algorithm computes the strongest paths between pairs of candidates by means of a variant of the Floyd Warshall  algorithm \cite{schulze2018schulze, csar2018computing}. Here the strength of a path between candidates is the strength of its lowest weight edge. Schulze then orders candidates by decreasing strength of their strongest paths. Our \fairSchulze \; method satisfies \mfracritbaserspace by producing the Schulze consensus ranking, ordering candidates by the strength of path - thereby optimizing for pairwise agreement, then correcting it to satisfy \mfracritfairspace with $\Delta$ desired fairness using \makefair. The algorithm’s pseudocode is provided in the supplement \cite{supplement}.

\noindent{\textbf{\fairSchulzespace Complexity.}}
\fairSchulzespace takes $O(n^2*|R|)$ to create the precedence matrix W, $O(n^3)$ to compute strongest paths to create a Schulze ranking \cite{schulze2018schulze, csar2018computing}, $O(nlogn)$ time to sort the candidates into correct order, and in the worst case $O(n^2*(|\mathcal{P}|+1) *n)$ to fairly reorder candidates.

\noindent{\textbf{\textit{Fair-Borda  Solution for MFCR.}}}
We create \fairBordaspace specifically as an \ourproblem solution for large consensus ranking problems
In a comparative study of Kemeny consensus ranking generation, Borda \cite{borda1784memoire} was shown to be the fastest Kemeny approximation method \cite{ali2012experiments}. Thus it is an ideal strategy
for tackling our \ourproblem problem.
More precisely, Borda \cite{borda1784memoire} is a positional rank aggregation function that ranks candidates in descending order based on a total number of points allotted to each candidate. The total points allotted to candidate $x_i$ correspond to the total number of candidates ranked lower than $x_i$ in all the base rankings. 
Our \fairBordaspace method utilizes Borda to efficiently aggregate the base rankings with minimal error by minimizing pairwise disagreement with the base rankings. Thus, it satisfies the \mfracritbaserspace criteria. Next, the \makefairspace subroutine is applied to  the resulting Borda ranking so that it  satisfies \mfracritfairspace with parameter $\Delta$. The algorithm’s pseudocode is found in the supplement \cite{supplement}.

\noindent\textbf{Complexity of \fairBorda.} 
\fairBordaspace takes $O(n*R)$ time to determine the points per candidates, $O(nlogn)$ time to order the candidates by total points, and  $O(n^2*(|\mathcal{P}|+1) *n)$ to fairly reorder candidates.

\subsection{Price of Fairness Measure for Fair Rank Aggregation}
%\ear{the term \mfracritfairspace is way too long - we need a short but catchy acronym}
Satisfying \mfracritfairspace incurs a toll in terms of \mfracritbaser. Intuitively, this toll is greatest when the base rankings $R$ have a low degree of fairness and the $\Delta$ parameter requires a high degree of fairness.
Thus, we now design the concept of  {\it Price of Fairness} ($PoF$) as a metric to quantify the cost of \mfracritfairspace as an increase in \pdmetricspace from the consensus ranking satisfying only \mfracritbaser.
We compute $PoF$ as the difference in the \pdmetricspace of the fair consensus ranking $\pi^{C*}$ and \pdmetricspace of the fairness unaware consensus ranking $\pi^{C}$:
%added for slides
\color{black}
\begin{equation}
    PoF =   PD\ Loss(R,\pi^{C*}) -  PD\ Loss(R,\pi^{C})\label{pricefairness}
\end{equation}
We note that $PoF$ is always $\geq 0$ as the fair consensus ranking \textit{at best} represents the preferences of the base rankings equally as well as the fairness-unaware consensus ranking.
As is true of the fairness and utility tradeoff in general \cite{singh2018fairness, yang2017measuring}, we note here and observe in our experiments that  $PoF$ for the \ourproblem problem can be significant.

\begin{table}[htbp]
\begin{center}
\caption{Mallows datasets: $\vert R \vert=150$ base rankings over $90$ candidates, with $6$ candidates in each of $15$ intersectional groups from $dom(Race)=5$ and $dom(Gender)=3$}

\begin{tabular}{|c|c|c|c|c|}
\hline
\textbf{Mallows Dataset}&\multicolumn{3}{|c|}{\textbf{Fairness metrics on modal ranking}} \\
\cline{2-4} 
 & \textbf{$ARP_{Gender}$}& \textbf{$ARP_{Race}$}& \textbf{$IRP$} \\
\hline
Low-Fair & $0.70$ & $0.70$ & $1.00$ \\
\hline
Medium-Fair & $0.50$ & $0.50$ & $0.75$ \\
\hline
High-Fair & $0.30$ & $0.30$ & $0.54$ \\
\hline
\end{tabular}
\label{MallowsDes}
\end{center}
\vspace{-3mm}
\end{table}

\begin{figure}[t]
     \centering
     \begin{subfigure}[t]{\columnwidth}
         \centering
         \includegraphics[width=.94\columnwidth]{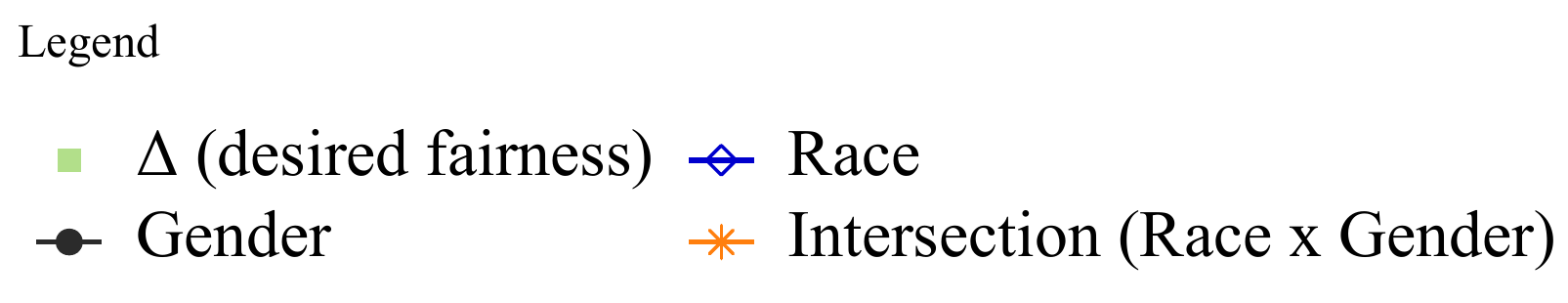}
         \label{legend}
     \end{subfigure}
     \vspace{.5em}% 
     \begin{subfigure}[t]{\columnwidth}
     \captionsetup{skip=1pt}
         \centering
         \includegraphics[width=.99\columnwidth]{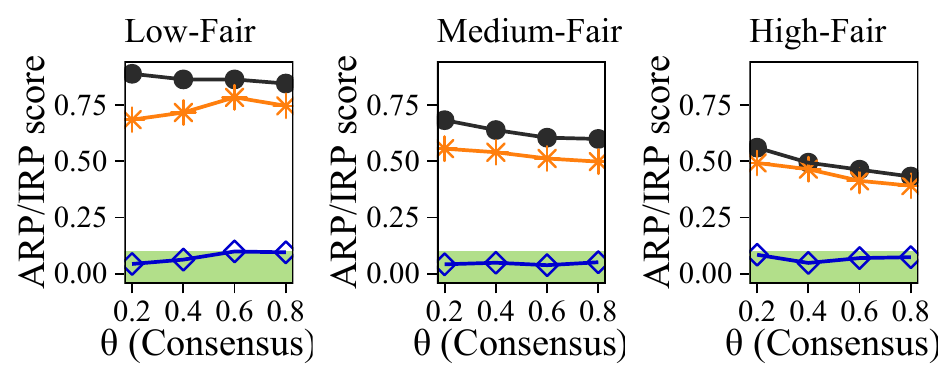}
     \end{subfigure}
     \hfill
     \vspace{.5em}% 
     \begin{subfigure}[t]{\columnwidth}
     \captionsetup{skip=1pt}
         \centering
         \includegraphics[width=.99\columnwidth]{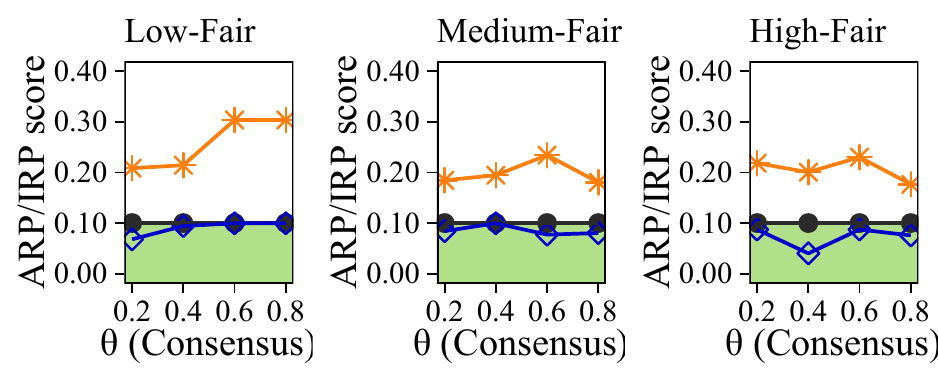}
         \caption{Protected attribute only group fairness approach}
     \end{subfigure}
     \hfill
     \vspace{.5em}% 
     \begin{subfigure}[t]{\columnwidth}
     \captionsetup{skip=1pt}
         \centering
         \includegraphics[width=.99\columnwidth]{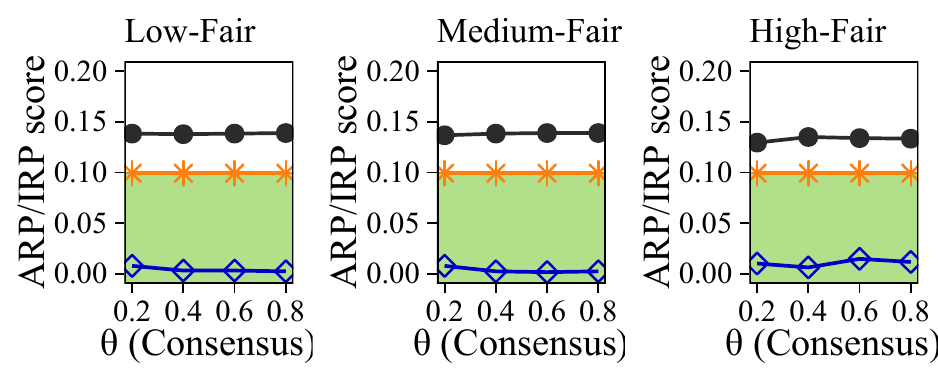}
         \caption{Intersection only group fairness approach}
     \end{subfigure}
     \hfill
     \vspace{.5em}% 
     \begin{subfigure}[t]{\columnwidth}
     \captionsetup{skip=1pt}
         \centering
         \includegraphics[width=.99\columnwidth]{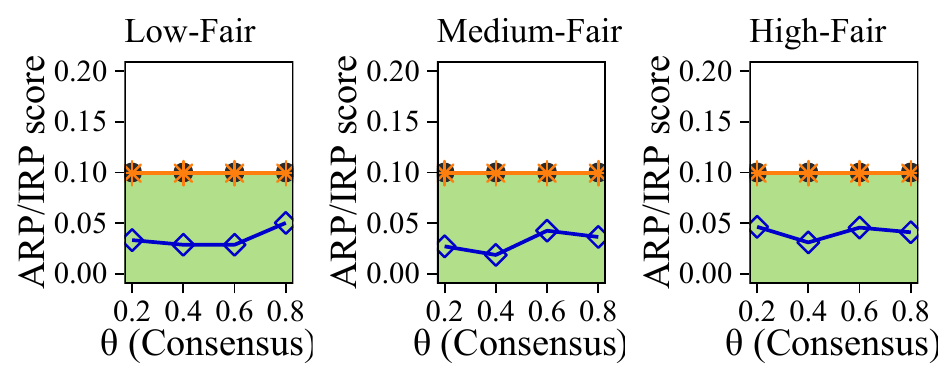}
         \caption{\fairnesscriteria group fairness approach in \fairKemeny}
     \end{subfigure}
     \hfill
        \caption{Comparing group fairness approaches for Mallows datasets}
        \label{fig:constraints}
\vspace{-4mm}
\end{figure}

\section{Experimental Study}\label{experiments}
\noindent\textbf{Experimental Methodology.}
We conduct a systematic study of alternate group fairness approaches 
% (including \fairnesscriterianospace) and 
evaluating our \ourproblem methods against  baselines under a rich variety of conditions in the base rankings 
modeled using the Mallows model \cite{mallows1957non,ali2012experiments}. 
In particular, we analyze how the degree of consensus and fairness present within the base rankings  along with the $\Delta$ parameter affect the PoF of the consensus ranking. 
We also study  the scalability of our \ourproblem solutions.
We conclude with a case study on student  rankings and merit scholarship \cite{kimmons}.

\begin{figure*}[t]
\centering{\includegraphics[width=.9\textwidth]{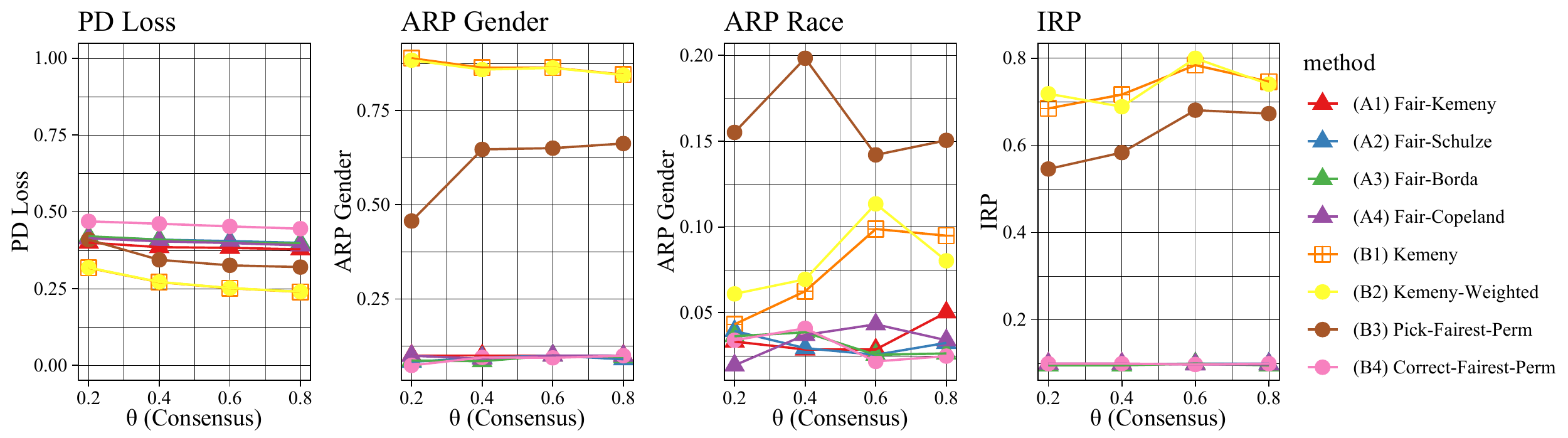}}
\caption{ Evaluating proposed \ourproblem methods: Low-Fair Dataset from Table $1$, with $\Delta = .1$ }
\label{fig:evalmethodspanel}
\vspace{-3mm}
\end{figure*}

\subsection{Studying Alternate Group Fairness Constraints}\label{constraint_analysis}
\color{black}
%\ear{Why is below blue. is it not basically unchanged?}
\noindent\textbf{Datasets.} We leverage the Mallows Model \cite{mallows1957non, ali2012experiments}, a probability distribution over rankings, as a data generation procedure.
Widely used to evaluate consensus ranking tasks \cite{kuhlman2019fare,ali2012experiments,brancotte2015rank}, the Mallows Model is an exponential location-spread model, in which the location is a modal ranking among a set of base rankings, denoted by $\pi'$, and the spread parameter, denoted by $\theta$, is a non-negative number.
For a ranking $\pi' \in S_n$, the Mallows model is the following probability distribution:
\begin{equation}
    P(\pi')_{\theta} = \frac{exp(-\theta * d(\pi', \pi))}{\psi(\theta)}\label{mallows}
\end{equation}
where $\psi(\theta)$ is the normalization factor that depends on the spread parameter $\theta$, and has a closed form.
Utilizing the Kendall-Tau distance as the distance metric $d(\pi', \pi)$, the Kemeny consensus ranking corresponds to the maximum likelihood estimator of the Mallows model \cite{young1995optimal}.
We control the fairness of  base rankings by setting the fairness in the modal ranking, consensus is adjusted via the $\theta$ parameter.
When $\theta = 0$ , there is no consensus among the base rankings around the modal ranking $\pi'$.
As $\theta$ increases, the base rankings gain consensus around the modal ranking $\pi'$.

\noindent\textbf{Experimentally Comparing 
\fairnesscriteria Criteria  with Alternate Group Fairness Criteria.}
%\ear{Same below. Why is below blue. is it not basically unchanged?}
To evaluate group fairness notions,
we compare three alternate group fairness approaches in rank aggregation with traditional Kenny rank aggregation
(Figure \ref{fig:constraints}).
As depicted in Figure \ref{fig:constraints}, 
we vary the spread parameter $\theta$ to create four sets of base rankings with different degrees of consensus for each modal ranking.

Group fairness strategies we compare include: only constraining the protected attributes $-$ our \fairKemenyspace with Equation \eqref{IRP} removed, only constraining the intersection $-$ our \fairKemenyspace with Equation \eqref{ARP} removed, and our proposed \fairKemeny. 
In all cases, we set our desired
fairness  threshold $\Delta = .1$  to specify close proximity to statistical parity as per  Definition 7.

%put paragraph break here
\color{black}
In Figure \ref{fig:constraints}, we observe that, under all fairness conditions and consensus scenarios ($\theta$), the Kemeny fairness unaware aggregation method creates a consensus ranking with $ARP/IRP$ scores significantly above the desired threshold.
The protected attribute-only approach consistently results in consensus rankings with Gender and Race   at or below the threshold.
But it still leaves $IRP$ significantly higher than desired.
The intersection only approach successfully constrains the intersection to the desired fairness level.
But it leaves the $ARP$ of Gender higher than $\Delta$. Our proposed \fairnesscriteria criteria is the only group fairness approach formulation which ensures that both the individual and intersection of  protected attributes are at or below the desired level of fairness. Thus, we conclude for a protected attribute or intersection to be \textit{guaranteed protected at a desired level of fairness it must be constrained explicitly}.

\subsection{Comparison of \ourproblem Solutions: Fairness and Preference Representation of Generated Consensus Ranking}

To analyze the efficacy of our proposed \ourproblem solutions in achieving \mfracritfairspace and \mfracritbaser, we compare our four methods against four baselines in Figure \ref{fig:evalmethodspanel}. Baselines are (1)
traditional \kem, (2) \weightedKemeny, which orders the base rankings from least  to most fair
and weights the fairest ranking by $|R|$ and the least fair by $1$, (3) \pickfairspace a variation of Pick-A-Perm \cite{schalekamp2009rank} which returns the fairest base ranking, and (4) \correctpermspace which utilizes \makefairspace to correct the fairest base ranking to satisfy $\Delta$.

Examining Figure \ref{fig:evalmethodspanel}, we see that \kemspace and \weightedKemenyspace perform best at representing the base rankings. \pickfairspace in the case the base rankings have a high degree of consensus ($\theta$) represents the base rankings as well. 
However, these methods do not achieve our fairness criteria, i.e., they perform the worst at satisfying \mfracritfair. While \kemspace does not attempt to incorporate group fairness, \pickfairspace and \weightedKemenyspace aim 
to incorporate fairness. They do not succeed as solutions to the \ourproblem problem -   because they do not achieve a desired level of fairness ($\Delta$ parameter). This results in a consensus ranking that at best represents the fairest ranking in the base set, i.e., \pickfairspace indeed  returns the fairest among the base rankings. Also, not surprisingly, \weightedKemenyspace is not fairer than Pick-Fairest-Perm ("fairer" defined as lower $IRP/ARP$ scores).
The last baseline Correct-Fairest-Perm  satisfies \mfracritfair (due to the utilization of \makefairalgo), but with a significantly higher \pdmetric. This indicates it does not represent the base rankings well, making it a poor \ourproblem solution.

Next, we examine our proposed methods. It is clear from the $ARP$ and $IRP$ graphs that all our methods achieve the desired level of fairness ($\Delta = 0.1$), thus satisfying \mfracritfair. When examining
\pdmetric, we see \fairKemenyspace performs best, which intuitively makes sense as it optimally minimizes pairwise disagreements subject to fairness constraints. Next, in order of decreasing
\pdmetric \; is \fairCopeland, then \fairSchulze, followed by  \fairBorda. This is also as expected,
as the first two methods are Condorcet methods, and \fairBordaspace is not. 
However, these polynomial-time algorithms compared to \fairKemenyspace perform comparably well in representing the base rankings -- this is particularly true when there is less consensus in the base rankings.

\subsection{Studying the Price of Fairness}

\begin{figure}[t]
\centering{\includegraphics[width=\columnwidth]{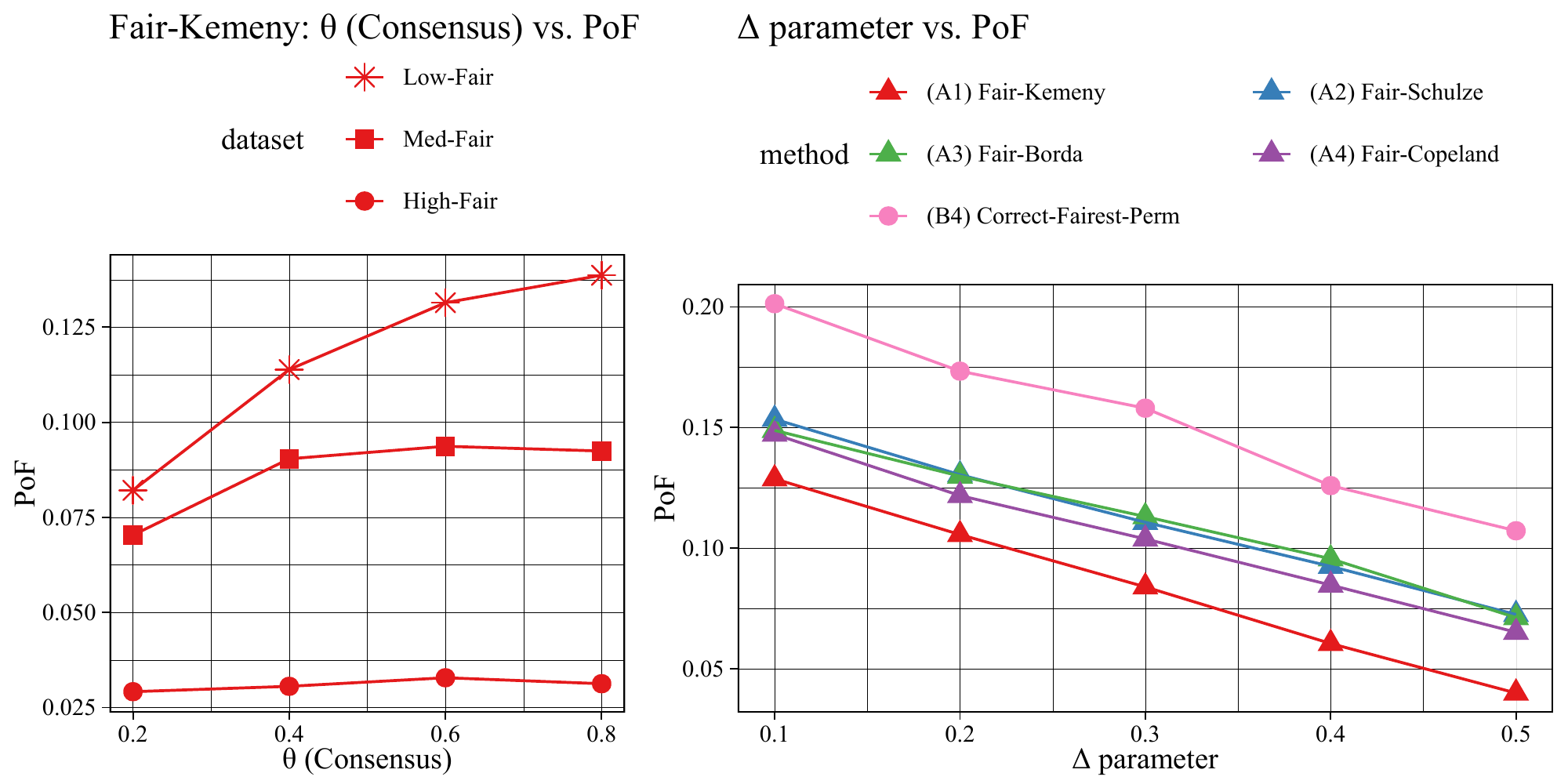}}
\caption{PoF Analysis: Datasets from TABLE \ref{MallowsDes}}
\label{fig:pof}
\vspace{-3mm}
\end{figure}

In Figure \ref{fig:pof}, we evaluate the Price of Fairness ($PoF$) using the metric from Equation \eqref{pricefairness}.
We analyze how the amount of consensus in the base rankings and the $\Delta$ parameter  affect  $PoF$.
Utilizing the \fairKemenyspace method, we observe that the fairness of the modal ranking has the biggest impact on  the price of fairness.
When the modal ranking has a higher level of fairness, the level of consensus around that ranking does not significantly impact the price of fairness.
But when the modal ranking has a very low level of fairness, the degree of consensus ($\theta$) has a larger impact.
A low degree of consensus has the effect of ``cancelling out" the fairness in the modal ranking. 
Intuitively, a high degree of consensus around a low fairness modal ranking results in a higher price of fairness.

Next, we examine the effect of the $\Delta$ parameter on  $PoF$. We uncover a steep inverse linear trend between $\Delta$ and the $PoF$ for  our four methods and Correct-Fairest-Perm on the Low-Fair dataset with $\theta = 0.6$. Across all methods that utilize the $\Delta$ parameter, when $\Delta$ is high,  $PoF$ is lower. This is intuitive as when $\Delta$ is small, the consensus ranking is likely required to be \textit{fairer than the base rankings}.
This  in turn increases the amount of disagreement between the consensus ranking and the base rankings.

\subsection{\ourproblem Solutions: Study of Scalability}\label{scalestudysect}

We evaluate  the scalability of all   methods  presented above. We have implemented our methods in python and used IBM CPLEX optimization software for Kemeny, \fairKemeny, and \weightedKemeny. All experiments were performed on a Windows 10 machine with $32$GB of RAM

\begin{figure}[t]
\centering{\includegraphics[width=.8\columnwidth]{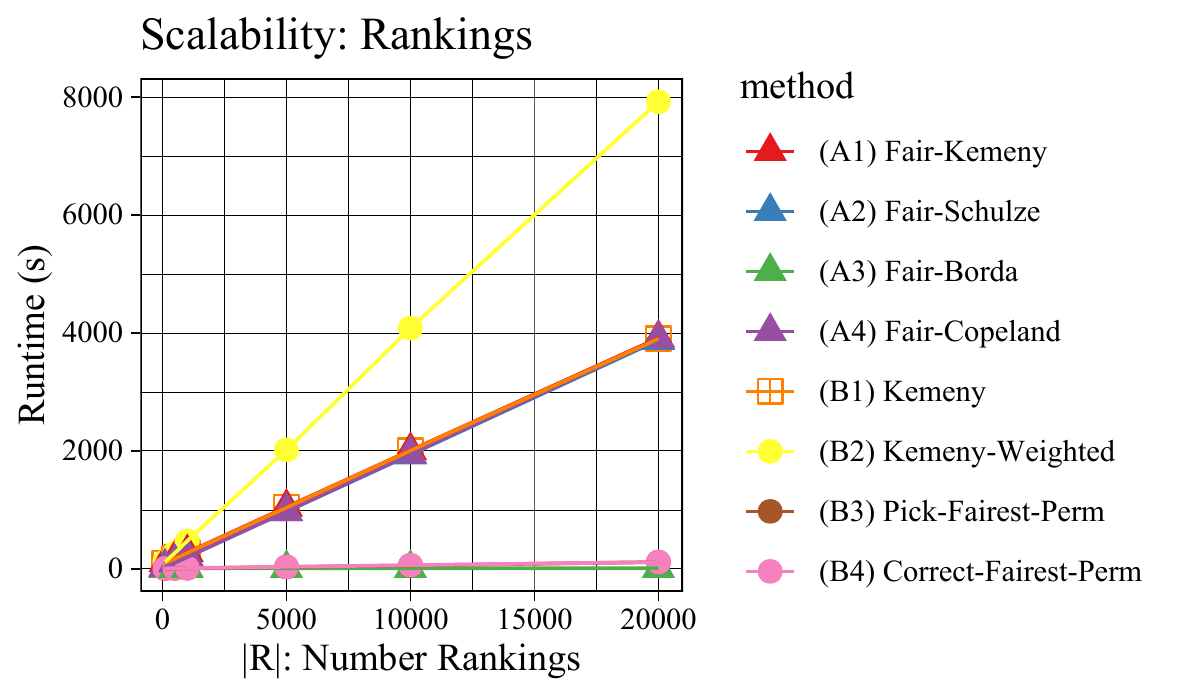}}
\caption{Scalability with an increasing number of base rankings}
\label{fig:rankings_Scale}
\vspace{-3mm}
\end{figure}

\noindent\textbf{Scalability in Number of Rankers.} In Figure \ref{fig:rankings_Scale}, we analyze the efficiency of our proposed methods in handling increasingly large numbers of base rankings. We create a Mallows model dataset with a modal ranking with
$ARP(Race) = 0.15,$  $ARP(Gender) = 0.7,  IRP =.55 $, $dom(Race) = 2$  and $dom(Gender) = 2$, we set ($\theta = .6)$, $n = 100$, and specify a tight fairness requirement with $\Delta = .1$. 

In Figure \ref{fig:rankings_Scale}, we see that three tiers of methods emerge. The fastest tier includes \fairBorda, \pickfair, and \correctperm, while the second is \fairSchulze, \fairCopeland, \fairKemeny, and Kemeny. We note that our proposed methods perform no slower than regular (not-fair) Kemeny. Lastly, \weightedKemenyspace performs the slowest due to having to order and weight large numbers of base rankings. 

Next, we study the scalability of the most efficient method proposed - \fairBordaspace on the same Mallows model dataset as above. In  Table \ref{FairBordaRankerScale}, we see that for \textit{tens of millions of rankings} on a Low-Fair modal ranking \fairBordaspace creates a fair consensus ranking \textit{in under a minute}.

\begin{figure}[t]
\centering{\includegraphics[width=.8\columnwidth]{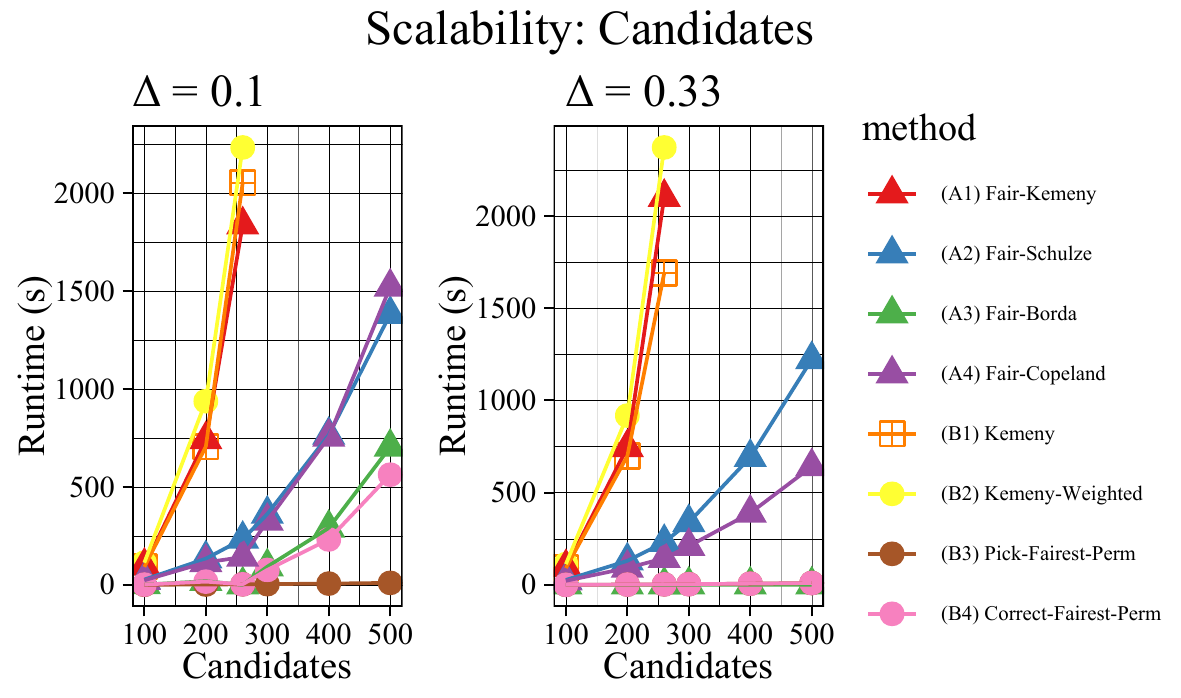}}
\caption{Scalability in increasing number of candidates}
\label{fig:candidate_Scale}
\end{figure}

\noindent\textbf{Scalability in Number of Candidates.} In contrast to base rankings, large numbers of candidates is a greater challenge for consensus generation. In Figure \ref{fig:candidate_Scale}, we analyze the efficiency of our methods for increasingly large numbers of candidates and the effect of the $\Delta$ parameter on the execution time.
We create a Mallows model dataset with a modal ranking with $ARP(Race) = 0.31 , ARP(Gender) = 0.44,  IRP =.45 $,$dom(Race) = 2$ and $dom(Gender) = 2$, we set ($\theta = .6)$, $|R| = 100$ and experiment with a tight fairness requirement with $\Delta = .1$ and a looser  but overall fairer than the base rankings $\Delta = .33$. In Figure \ref{fig:candidate_Scale}, we see the same tiers of methods as in Figure \ref{fig:rankings_Scale}. The optimization methods are the slowest and constrained by CPLEX's utilization of the machine's memory. Though we again note that our \fairKemenyspace is comparable to Kemeny and both are faster than \weightedKemeny. The optimization methods upper bound the polynomial time ones in order of decreasing execution time from  \fairSchulze, \fairCopeland, to \fairBorda. \fairBordaspace performs the fastest  comparable to the inferior \correctpermspace and  \mfracritfairspace Pick-Fairest-Perm. We see that a higher $\Delta$ parameter intuitively decreases the execution time. 

Lastly, in Table \ref{FairBordaCandidateScale}, we study the scalability of the most efficient method proposed - \fairBordaspace on the same dataset as the candidate study above.
For  $\Delta = .33$, we see that for tens of thousands of candidates, \fairBordaspace creates a fair consensus ranking in a handful of minutes.

\begin{table}[t]
%\color{blue}
\caption{\fairBordaspace Ranker Scale}
\vspace{-3mm}
\begin{center}
\begin{tabular}{ |c|c| } 
\hline
 \textbf{$|R|$ Number of Rankings} & \textbf{Execution time (s)}\\
 \hline
 $1,000$ & $4.8$ \\
 \hline
 $10,000$ & $4.81$\\ 
 \hline
 $100,000$ & $5.21$\\ 
 \hline
 $1,000,000$ & $9.36$\\ 
 \hline
 $10,000,000$ & $50.75$\\ 
 \hline
\end{tabular}
\end{center}
\label{FairBordaRankerScale}
\vspace{-7mm}
\end{table}

\begin{table}[t]
%\color{blue}
\caption{\fairBordaspace Candidate Scale}
\vspace{-3mm}
\begin{center}
\begin{tabular}{ |c|c| } 
\hline
 \textbf{$|X|$ Number of Candidates} & \textbf{Execution time (s)}\\
 \hline
 $1,000$ & $0.37$ \\
 \hline
 $10,000$ & $30.83$\\ 
 \hline
 $20,000$ & $121.49$\\ 
 \hline
 $30,000$ & $273.24$\\ 
 \hline
 $40,000$ & $482.29$\\ 
 \hline
 $50,000$ & $749.00$\\ 
 \hline
 $100,000$ & $3007.19$\\ 
 \hline
\end{tabular}
\end{center}
\label{FairBordaCandidateScale}
\vspace{-6mm}
\end{table}

\subsection{Empirical Takeaways}
When utilizing our \ourproblem methods, we recommend \fairKemenyspace for smaller candidate databases and note that the number of rankings can be reasonably large (thousands). Next, \fairCopelandspace and \fairSchulzespace provide nearly comparable performance on larger candidate databases and number of rankings when  fairness requirement is strict. However, if fairness requirement is looser, we recommend \fairCopelandspace for more efficiency with decreased $PoF$. If the consensus ranking problem is very large, \fairBordaspace is the best choice with significant speed-up and minimal increase in $PoF$.

\color{black}
\subsection{Case Study of Student Merit Scholarships}

\begin{table*}
%\color{blue}
\caption{\textbf{Exam Case Study:} Attribute values columns (e.g, Men, SubLunch) indicate $FPR$ scores, and $Gender$, $Race$, and $Lunch$ indicate $ARP$ scores.}
\begin{center}
\begin{tabular*}{\textwidth}{|c|c|c|c|c|c|c|c|c|c|c|c|c|c|}

 \hline
 \textbf{\textit{Ranking}} & \textbf{\textit{Men}} & \textbf{\textit{Women}} & \textbf{\textit{Gender}} & \textbf{\textit{NoSub}} & \textbf{\textit{SubLunch}}  & \textbf{\textit{Lunch}} & \textbf{\textit{Asian}} & \textbf{\textit{White}} & \textbf{\textit{Black}} & \textbf{\textit{AlaskaNat}}. & \textbf{\textit{NatHaw.}} & \textbf{\textit{Race}} & \textbf{\textit{IRP}} \\ [0.5ex] 
 \hline
 \textit{Math} & 0.39 & 0.61 & 0.37 & 0.72 & 0.28 & 0.44 & 0.59 & 0.49 & 0.56 & 0.55 & 0.22 & 0.37 & 0.65\\
 \hline
 \textit{Reading} & 0.60  & 0.4 & 0.20 & 0.63 & 0.37 & 0.26 & 0.55 & 0.45 & 0.56 & 0.56 & 0.29 & 0.27 & 0.47\\
 \hline
 \textit{Writing} & 0.63 & 0.37 & 0.24 & 0.68 & 0.32 & 0.36 & 0.56 & 0.47 & 0.56 & 0.52 & 0.32 & 0.24 & 0.51\\
 \hline
 Kemeny & 0.57 & 0.43 & 0.14 & 0.67 & 0.33 & 0.34 & 0.57 & 0.46 & 0.57 & 0.54 & 0.27 & 0.30 & 0.52\\
 \hline
 \textbf{\fairKemeny} & 0.51 & 0.49 & \textbf{0.02} & 0.52 & 0.48 & {\bf0.04} & 0.52 & 0.50 & 0.51 & 0.50 & 0.47 & {\bf0.04} & {\bf0.05}\\
 \hline
 \textbf{\fairSchulze} & 0.50 & 0.50 & {\bf0.0} & 0.52 & 0.48 & {\bf0.04} & 0.52 & 0.50 & 0.50 & 0.51 & 0.47 & {\bf0.05} & {\bf0.05}\\
 \hline
 \textbf{\fairBorda} & 0.50 & 0.50 & {\bf0.0} & 0.52 & 0.48 & {\bf0.04} & 0.52 & 0.50 & 0.51 & 0.50 & 0.48  & {\bf0.04} & {\bf0.05}\\
 \hline
 \textbf{\fairCopeland} & 0.51 & 0.49 & {\bf0.02} & 0.52 & 0.48 & {\bf0.04} & 0.52 & 0.49 & 0.51 & 0.50 & 0.48 & {\bf0.04} & {\bf0.05}\\
 \hline

 \hline

\end{tabular*}
\end{center}
\label{exam_casestudy}
\vspace{-7mm}
\end{table*}

We demonstrate that our \ourproblem solutions create real-world fair consensus rankings over students with multiple protected attributes.
Entrance exam and test scores are commonly used as part of admissions decisions or merit scholarship allocations for educational institutions ranging from magnet schools\cite{abdulkadirouglu2014elite} to graduate admission \cite{bleske2014trends}.
As sociodemographic factors such as student socioeconomic status, race, and gender can have a large effect on exam outcomes \cite{reiss2013socioeconomic, fisk2018s, salehi2019gender}, schools and testing organizations are exploring ways to level the playing field when using exam scores for admission decisions\cite{jaschik2019new}.

We utilized a publicly available dataset \cite{kimmons} modeling student exam scores for math, reading and writing exams.
The data contained three protected attributes of {\tt Gender}  (man or woman), {\tt Race}  (5 racial groups) and {\tt Lunch}  (if student received reduced cost lunches).
We utilized the exam scores provided in each subject to create $\vert R \vert = 3$ base rankings (ordered by score) over $200$ students.

In Table \ref{exam_casestudy},  all protected attributes have an $ARP \geq .2$ (with higher $IRP$ scores) across all base rankings -- indicating statistical parity is far from being achieved. This contrast is particularly stark as we can see students with subsidized lunches are ranked low, along with NatHawaiin students. Also, there appears to be a substantial gender imbalance.
We create a (fairness-unaware) Kemeny consensus ranking, and observe that the biases in the base rankings are unfortunately also reflected in the consensus ranking.
Thus, if the consensus ranking was used to determine merit scholarships then the students who receive subsidized lunch would receive almost \textit{three times less aid} as the group of students who do not require subsidized lunches. The group of men would also receive more merit aid then women. NatHawaiian students would received almost half the amount of aid as Asian and Black students.

We then compare this to utilizing our four proposed \ourproblem solutions,
employing the \fairnesscriteria criteria and setting $\Delta = .05$ to ensure almost perfect statistical parity across all protected attributes and their intersection.
All methods generate a de-biased consensus ranking, with $ARP \leq .05$ and $IRP \leq .05$. This translates to all groups receiving an extremely close to equal proportion of merit scholarships. The disparities between merit aid received by the men and women are nearly nonexistent, and the difference between racial groups is leveled. The severe advantage of students who do not require subsidized lunch is also removed. As conclusion, utilizing the fair consensus rankings created by our \ourproblem solutions ensures certain student groups are not disadvantaged in the merit aid process. 

\color{black}

\section{Related Work}

\noindent\textbf{Fair Ranking Measures.} While the majority  of  algorithmic fairness work has concentrated on the task of binary classification, several notions of group fairness have been defined for (single) rankings.
The most widely adopted notion of group fairness in rankings is statistical parity
\cite{yang2017measuring, kuhlman2019fare, geyik2019fairness, beutel2019fairness, narasimhan2020pairwise, zehlike2017fa, kuhlman2020rank, hertweck2021moral}.
Most works focus on measuring and enforcing statistical parity between two groups defined by a single binary protected attribute \cite{kuhlman2019fare, singh2018fairness, yang2017measuring, zehlike2017fa, beutel2019fairness}. 
The pairwise fairness metrics for a single protected attribute of binary groups of \cite{kuhlman2019fare} inspire our metrics, we propose metrics extending the pairwise approach to multi-valued attributes and for multiple attributes.

The recent works of Narasimhan et al. \cite{narasimhan2020pairwise} and Geyik et al. \cite{geyik2019fairness} propose group fairness metrics for a single multi-valued protected attribute of multiple values.
Narasimhan et al. \cite{narasimhan2020pairwise} introduce a pairwise accuracy statistical parity notion for ranking and regression models.
Our pairwise statistical parity notion differs by counting pairs directly as consensus ranking is not performed from a model whose accuracy we can constrain. 
Geyik et al. \cite{geyik2019fairness} formulate group fairness by casting the group positive outcome as inclusion in the top-$k$.
As consensus generation combines multiple whole rankings, simply setting $k = n$ would not capture group fairness for a consensus ranking.  

\noindent\textbf{Multiple Protected Attributes.} 
Recent work addressing group fairness in multiple protected attribute settings is entirely focused on the binary classification task. 
Kearns et al. \cite{kearns2018preventing} introduced procedures for auditing and constructing classifiers subject to group fairness notions where groups could be defined as combinations of protected attributes.
Hebert-Johnson at al. \cite{hebert2018multicalibration} proposed 
% multi-calibrated classifier, 
an approach which ensures accurate predictions on all combinations of groups formed from protected attributes.
Foulds et al. \cite{foulds2020intersectional} proposed differential fairness; an intersectional fairness notion, for ensuring group fairness with respect to classification outcome probabilities for all possible combinations of sub-groups while ensuring differential privacy.

Within the domain of rankings, Yang et al. \cite{yang2019balanced} design algorithms to mitigate within-group unfairness in the presence of diversity requirements for groups encoded by multiple protected attributes $-$ we consider fairness between groups. 
Yang et al. \cite{yang2020causal} present a counterfactual causal modelling approach to create intersectionally fair rankings in score-based and learning-to-rank models.
While we assume access to the base rankings, we do not know why they reflect a specific order or how the rankings would differ based on changes in the protected attributes of candidates.
Without these counterfactual outcomes, causal fairness notions are difficult to deploy.
 
\noindent\textbf{Rank Aggregation.} Rank aggregation originates from the study of ranked ballot voting methods in Social Choice Theory \cite{arrow2012social, kemeny1959mathematics, brandt2012computational}.
Rank aggregation aims to find the consensus ranking which is the closest to the set of base rankings.
This has been studied in information retrieval \cite{dwork2001rank}, machine learning \cite{klementiev2008unsupervised}, databases \cite{brancotte2015rank} and voting theory \cite{mao2012social}.
Kemeny rank aggregation, a preeminent rank aggregation method satisfying several social choice axioms \cite{brandt2012computational},
has been applied to a wide set of applications: MRNA gene rankings \cite{madarash2004enhancing}, teacher evaluations \cite{bury2003application}, and conference paper selections \cite{baskin2009preference}.
Recent work
%% The work of Kuhlman and Rundensteiner 
\cite{kuhlman2020rank}, which introduced the fair rank aggregation problem, also leverages  Kemeny rank aggregation. However, they  assume a single binary protected attribute. 
Our work instead 
% advances fair rank aggregation
now handles
% by addressing how to create a consensus ranking that is fair with respect to a 
multi-valued as well as multiple protected attributes.    

\section{Conclusion}

This work introduces the first solution to the  multi-attribute fair consensus ranking (MFCR) problem. 
First, we design novel \fairnesscriteria fairness criteria to support  interpretable tuning of fair outcomes for groups defined by multiple protected attributes in consensus rankings.
We then design four alternate \ourproblem algorithms using our
proposed \fairnesscriteria model.
We demonstrate the efficacy, scalability, and quantify the price of fairness achieved by our MFCR solutions in selecting a fair consensus ranking over a vast array of rank aggregation scenarios. Code, metrics, supplemental material, and experiments made publicly available. \url{https://github.com/KCachel/MANI-Rank}

\color{black}
\newpage

\bibliographystyle{IEEEtran}
\bibliography{main}

\appendix

\section{Case Study of Computer Science Department Rankings}

\begin{table*}
\caption{\textbf{CSRankings Study:} Attribute values columns (e.g, Northeast, Prive) indicate $FPR$ scores, and Location, and Type indicate $ARP$ scores.}
\centering
\resizebox{\textwidth}{!}{%
\begin{tabular}{|l|l|l|l|l|l|l|l|l|l|}
\hline
\textit{\textbf{Ranking}} &
  \textit{\textbf{Northeast}} &
  \textit{\textbf{Midwest}} &
  \textit{\textbf{West}} &
  \textit{\textbf{South}} &
  \textit{\textbf{Location}} &
  \textit{\textbf{Private}} &
  \textit{\textbf{Public}} &
  \textit{\textbf{Type}} &
  \textit{\textbf{IRP}} \\ \hline
2000          & 0.692 & 0.373 & 0.574 & 0.315 & 0.377 & 0.642 & 0.358 & 0.284 & 0.536 \\ \hline
2001          & 0.712 & 0.370 & 0.598 & 0.270 & 0.442 & 0.608 & 0.392 & 0.217 & 0.582 \\ \hline
2002          & 0.700 & 0.451 & 0.610 & 0.203 & 0.497 & 0.569 & 0.431 & 0.138 & 0.574 \\ \hline
2003          & 0.689 & 0.392 & 0.589 & 0.287 & 0.402 & 0.562 & 0.438 & 0.124 & 0.557 \\ \hline
2004          & 0.658 & 0.481 & 0.570 & 0.265 & 0.392 & 0.610 & 0.390 & 0.221 & 0.520 \\ \hline
2005          & 0.680 & 0.457 & 0.552 & 0.278 & 0.402 & 0.607 & 0.393 & 0.213 & 0.480 \\ \hline
2006          & 0.707 & 0.451 & 0.550 & 0.255 & 0.452 & 0.582 & 0.418 & 0.164 & 0.476 \\ \hline
2007          & 0.691 & 0.435 & 0.601 & 0.236 & 0.455 & 0.607 & 0.393 & 0.213 & 0.553 \\ \hline
2008          & 0.724 & 0.416 & 0.550 & 0.265 & 0.459 & 0.610 & 0.390 & 0.221 & 0.515 \\ \hline
2009          & 0.693 & 0.469 & 0.599 & 0.205 & 0.488 & 0.591 & 0.409 & 0.181 & 0.570 \\ \hline
2010          & 0.687 & 0.426 & 0.582 & 0.268 & 0.419 & 0.608 & 0.392 & 0.217 & 0.521 \\ \hline
2011          & 0.714 & 0.426 & 0.573 & 0.245 & 0.470 & 0.607 & 0.393 & 0.215 & 0.559 \\ \hline
2012          & 0.694 & 0.460 & 0.574 & 0.237 & 0.457 & 0.582 & 0.418 & 0.164 & 0.542 \\ \hline
2013          & 0.738 & 0.436 & 0.533 & 0.249 & 0.489 & 0.606 & 0.394 & 0.211 & 0.545 \\ \hline
2014          & 0.674 & 0.476 & 0.561 & 0.259 & 0.416 & 0.582 & 0.418 & 0.164 & 0.495 \\ \hline
2015          & 0.726 & 0.469 & 0.532 & 0.236 & 0.490 & 0.599 & 0.401 & 0.197 & 0.561 \\ \hline
2016          & 0.711 & 0.470 & 0.517 & 0.267 & 0.445 & 0.606 & 0.394 & 0.211 & 0.512 \\ \hline
2017          & 0.688 & 0.478 & 0.540 & 0.264 & 0.424 & 0.547 & 0.453 & 0.095 & 0.448 \\ \hline
2018          & 0.661 & 0.504 & 0.559 & 0.253 & 0.409 & 0.551 & 0.449 & 0.103 & 0.457 \\ \hline
2019          & 0.680 & 0.499 & 0.531 & 0.264 & 0.416 & 0.566 & 0.434 & 0.132 & 0.467 \\ \hline
2020          & 0.666 & 0.445 & 0.540 & 0.318 & 0.348 & 0.579 & 0.421 & 0.158 & 0.415 \\ \hline
Kemeny        & 0.712 & 0.445 & 0.570 & 0.233 & 0.479 & 0.602 & 0.398 & 0.203 & 0.570 \\ \hline
Fair-Kemeny   & 0.532 & 0.499 & 0.532 & 0.432 & 0.100 & 0.498 & 0.502 & 0.004 & 0.099 \\ \hline
Fair-Schulze  & 0.529 & 0.501 & 0.529 & 0.436 & 0.093 & 0.494 & 0.506 & 0.012 & 0.089 \\ \hline
Fair-Borda    & 0.519 & 0.499 & 0.534 & 0.445 & 0.089 & 0.496 & 0.504 & 0.008 & 0.094 \\ \hline
Fair-Copeland & 0.531 & 0.501 & 0.531 & 0.432 & 0.099 & 0.494 & 0.506 & 0.012 & 0.099 \\ \hline
\end{tabular}%
}
\label{tab.csrankings}
\end{table*}

While group fairness concerns typically emerge when people are being being ranked, they also apply for other ranked entities.
We illustrate the effectiveness our framework on removing bias in Computer Science Department rankings using the publicly available CSRankings from csrankings.org \cite{berger2018csrankings}.
We collected rankings of $65$ departments in the US over the period of $2000-2020$ (utilizing the relative order presented on csrankings.org as the ranking for each year) to generate a 20-year consensus ranking.
To examine geographic and private vs. public institutional bias, we used the location (Northeast, South, West, MidWest) and type (Public or Private) of each institution as protected attributes.

In Table \ref{tab.csrankings}, we observe that the rankings over the years indeed exhibit bias, namely, a strong bias in the location attribute (high $ARP$) - which stems from institutions in the Northeast region commonly appearing at the top of the rankings, along with those in South region exhibiting the opposite trend. Additionally, there is a significant amount of intersectional bias $-$ resulting from the base rankings having Private and Northeast colleges highly ranked.

When using fairness-unaware Kemeny to create the 20-year consensus ranking, we observe that this bias is amplified, with the Location $ARP$ score resembling that of base rankings with higher such scores, and the $IRP$ score close to $0.6$.
By utilizing our \fairnesscriteria criteria and setting the group fairness threshold $\Delta = .05$, it can be seen that we were able to remove the bias in the consensus ranking.
All our proposed methods remove the bias toward Northeast and Private universities.

\end{document}